\documentclass[usegraphix,usenatbib]{mn2e}

\usepackage{amsmath}
\usepackage{amssymb}
\usepackage{times}
\usepackage{graphicx}
\usepackage{subfigure}
\usepackage{epsfig}
\usepackage{txfonts}
\usepackage{calc}

\def\apjs{{ApJS}}

\def\aap{{AA}}

\begin{document}

\pagenumbering{arabic}

\title[Models of Bars I: Early Types]{Models of Bars I:
  Flattish Profiles for Early-Type Galaxies}

\author[Williams \& Evans]
  {A.~A. Williams$^1$\thanks{E-mail:aamw3, nwe@ast.cam.ac.uk},
   N.~W. Evans$^1$
 \medskip
 \\$^1$Institute of Astronomy, University of Cambridge, Madingley Road,
       Cambridge, CB3 0HA, UK}
\maketitle

\begin{abstract}
We introduce a simple family of barred galaxy models with flat
rotation curves. They are built by convolving the axisymmetric
logarithmic model with a needle density.  The density contours in the
bar region are highly triaxial and elongated, but become spheroidal in
the outer parts. The mass distribution differs markedly from the
elliptical shape assumed in other analytical models, like Ferrers or
Freeman bars. The surface density profile along the bar major axis is
flattish, as befits models for bars in early-type galaxies (SB0,
SBa). The two-dimensional orbital structure of the models is analyzed
with surfaces of section and characteristic diagrams and it reveals
qualitatively new features. For some pattern speeds, additional
Lagrange points can exist along the major axis, and give rise to
off-centered, trapped orbits.  When the bar is weak, the orbital
structure is very simple, comprising just prograde, aligned $x_1$
orbits and retrograde anti-aligned $x_4$ orbits. As the bar strength
increases, the $x_1$ family becomes unstable and vanishes, with
propeller orbits dominating the characteristic diagram.
\end{abstract}


\section{INTRODUCTION}

A substantial fraction of all disc galaxies are barred. For example,
the {\tt Galaxy Zoo} project examined a volume-limited sample of 13
665 local disc galaxies and find the barred fraction to be $29.4 \pm
0.5$ per cent~\citep{Ma11}. It has been known since \citet{Va64} that
the Milky Way is barred, as confirmed most clearly in the
near-infrared images of the Galactic bulge seen by the COBE satellite
~\citep{Dw95}. The barred nature of the Andromeda Galaxy (M31) was
known still earlier, as \citet{Li56} first suggested this based on the
shapes of the inner isophotes.

Nonetheless, models of bars -- both for photometric fitting and for
dynamical investigations -- remain rather primitive. In many
studies~\citep[see e.g.,][]{Va72,Pa83,Te85,At83,Pf84}, the bar models
have density law
\begin{equation}
\rho(x,y,z) = \begin{cases}\rho_0 (1-m^2)^n & m<1 \\
                           0 & m\geq 1, \end{cases}
\label{eq:ferrers}
\end{equation}
with $m^2 = x^2/a^2 + y^2/b^2 + z^2/c^2$. Here $a,b$ and $c$ are the
constant semiaxes of the ellisoidal density contours, whilst $n$ is an
integer. The advantage of this model is that the gravitational
potential within the bar is a simple polynomial of order $2n+2$ in
$x,y$ and $z$. Exterior to the bar, the coefficients of the polynomial
now also depend on the radial confocal elliptic coordinate, and so the
potential is more complicated.  These expressions were originally
derived by nineteenth century mathematicians \citep[][see also Binney
  \& Tremaine 1987]{Fe77}, though \citet{Pf84} provided a simple and
efficient numerical algorithm for computing the potential. When $n=0$,
the models are the homogeneous ellipsoids investigated by
\citet{Fr66a,Fr66b}. The gravitational potential is now a quadratic
function of the spatial coordinates, and so the stellar orbits within
the bar are integrable.  However, the disadvantage of all these models
is that they are not particularly realistic. They possess more
homogeneous density profiles than typically observed.

In theoretical investigations, it is sometimes preferable to start
with a simple potential, rather than a density. For barred galaxies,
the potential can always be written as a Fourier series, and the $m=2$
azimuthal harmonic alone retained to give simple barred potentials. A
model that was introduced by \citet{Ba67} is
\begin{equation} 
\Phi(r, \theta) = V_0(r,\theta) + \epsilon \sqrt{r} (r_{\rm end}-r)\cos 2\phi,
\end{equation}
where ($r,\theta,\phi$) are spherical polar coordinates, $\epsilon$ is
small, $r_{\rm end}$ marks the end of the bar and $V_0(r,\theta$) is
an axisymmetric model of the Galaxy. This model is not without its
shortcomings, as the acceleration is not well-defined at the centre,
and the radial force is repulsive at $r \approx r_{\rm
  end}$. Nonetheless, with $V_0$ chosen as the isochrone, the orbital
structure of this model as a function of bar strength $\epsilon$ has
been discussed extensively in \citet[][and references
  therein]{CO79,CO80}.

Observational evidence from the surface photometry of barred galaxies
suggests that there exists a dichotomy~\citep{El85,Se93}. In
early-type galaxies (SB0, SBa), the surface brightness falls slowly
along the bar major axis. Very occasionally, the surface brightness is
nearly constant to the end of the bar. In late-type galaxies (SBb,
SBc), however, the light profiles are strongly falling exponentials
along the major axis. For example, \citet{El96} took $J$, $H$ and $K$
infrared band observations for a sample of barred glaxies across all
Hubble types. They confirmed that early-types have flattish light
profiles, whilst late-types have exponential profiles. The flattish
profiles arise from an abundance of old and young stars at the ends of
the bar.

Clearly, there is ample scope for the development of
  new models of barred galaxies.  This is the first of two papers in
  which we provide models for the two regimes -- flattish profiles and
  exponential profiles. In both cases, we exploit an ingenious
  algorithm introduced by \citet{Lo92}, which proceeeds by convolving
  an underlying spherical or axisymmetric potential with a needle-like
  density to give barred or triaxial models.

  Section 2 introduces our model with a flattish light profile
  suitable for early-types, which we call the {\it logarithmic bar}. It
  possesses an asymptotically flat rotation curve, so the name is
  doubly appropriate. The inner parts are strongly triaxial, whilst
  the outer parts are axisymmetric. Unlike Ferrers bars, our model
  does not need to be supplemented by other components representing
  the Galactic disk and halo~\citep[e.g.,][]{Te85,Ka05}.  Section 3
  discusses the orbital structure in the weak and strong bar
  r\'egimes. The weak bars are dominated by prograde $x_1$ and
  retrograde $x_4$ orbits, much like embedded Ferrers bars with low
  figure rotation~\citep[e.g.,][]{Te85}. The strong bars with high
  pattern speed exhibit a number of interesting dynamical features,
  including secondary Lagrange points on the major axis around which
  off-centered $x_4$ orbits librate.  Successive bifurcations of the
  $x_1$ family now yields an overwhelming preponderance of very
  slender, aligned orbits similar to propeller orbits~\citep{Ka05}.
  The rapidly rotating logarithmic bars therefore provide a counter-example
  to the notion that bars must be primarily supported by the $x_1$
  family. The propellers can be made extremely narrow -- unlike true
  $x_1$ orbits -- and so they may support highly elongated needle-like
  bars. Finally, section 4 summarises the principal properties of our
  new bar model.

\begin{figure}
\begin{centering}
  \includegraphics[scale=0.1]{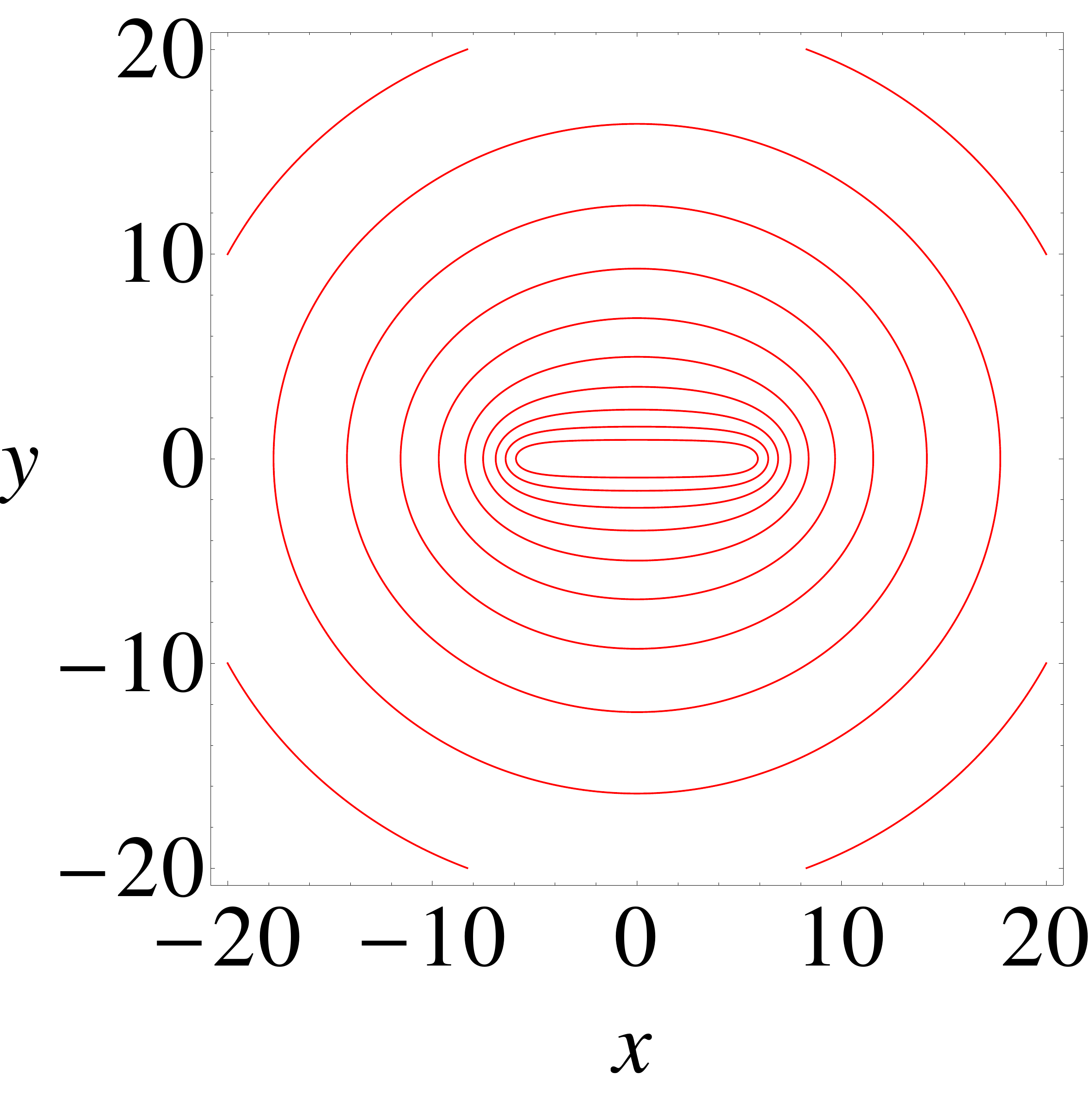}\quad\includegraphics[scale=0.1]{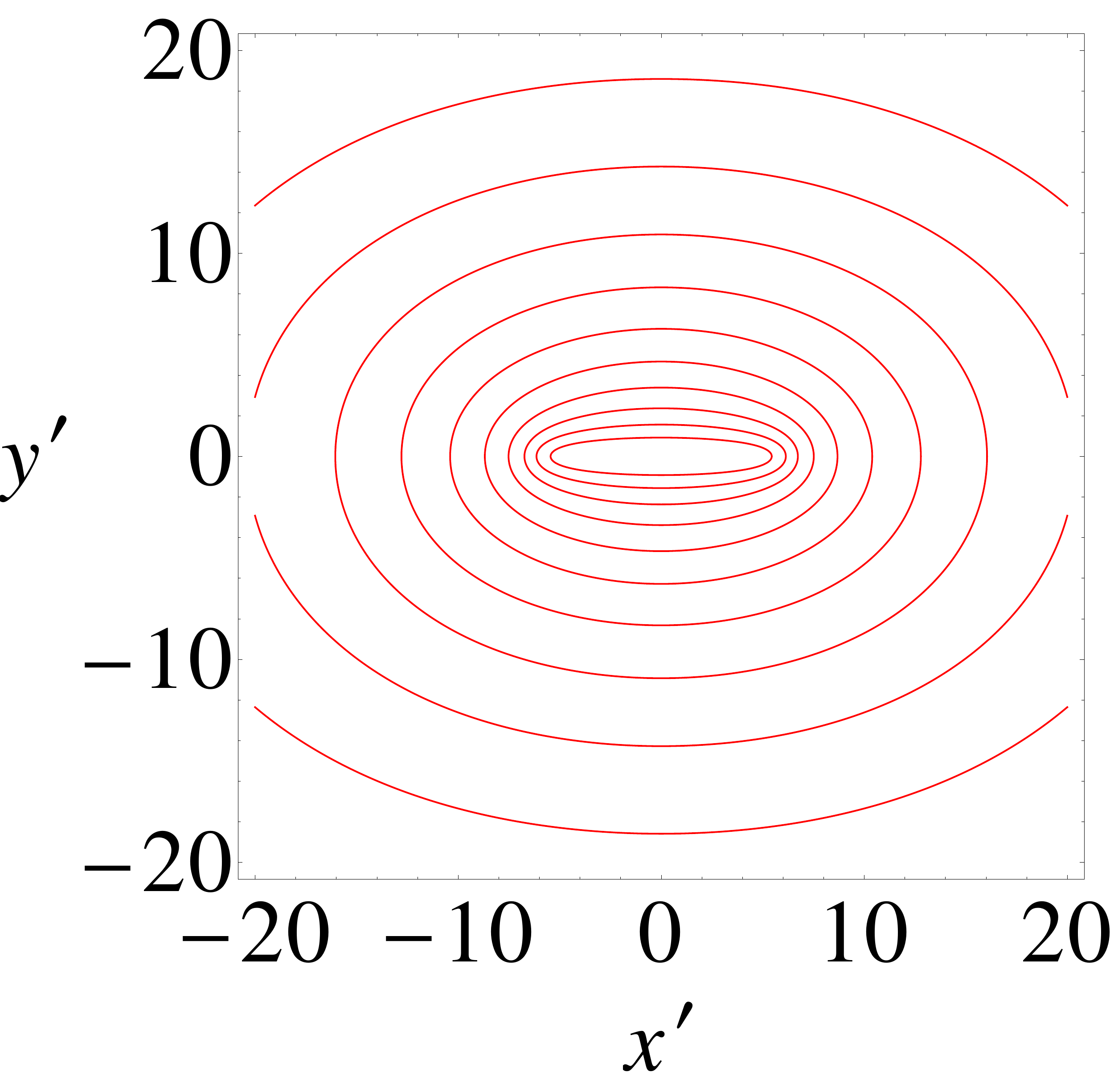}\quad\includegraphics[scale=0.1]{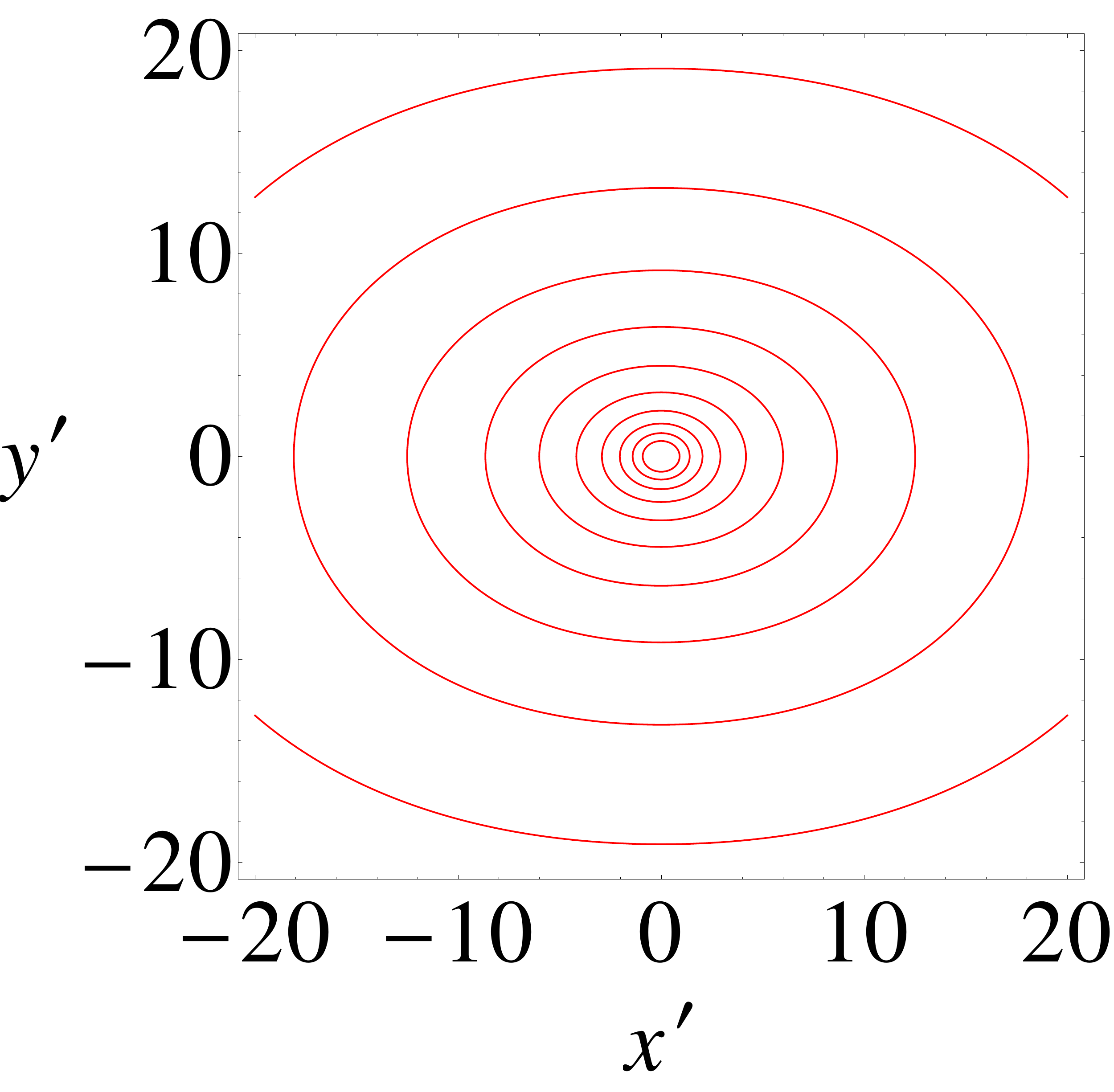}
  \caption{Logarithmically spaced contour plots of density in the
    three principal planes for a model with $v_{0}=1$, $q=0.9$ and
    $a=6$. Distances are given in units of $R_{\rm c}$.}
  \par\end{centering}
\label{fig:dens}
\end{figure}
\begin{figure}
\begin{centering}
\includegraphics[scale=0.1]{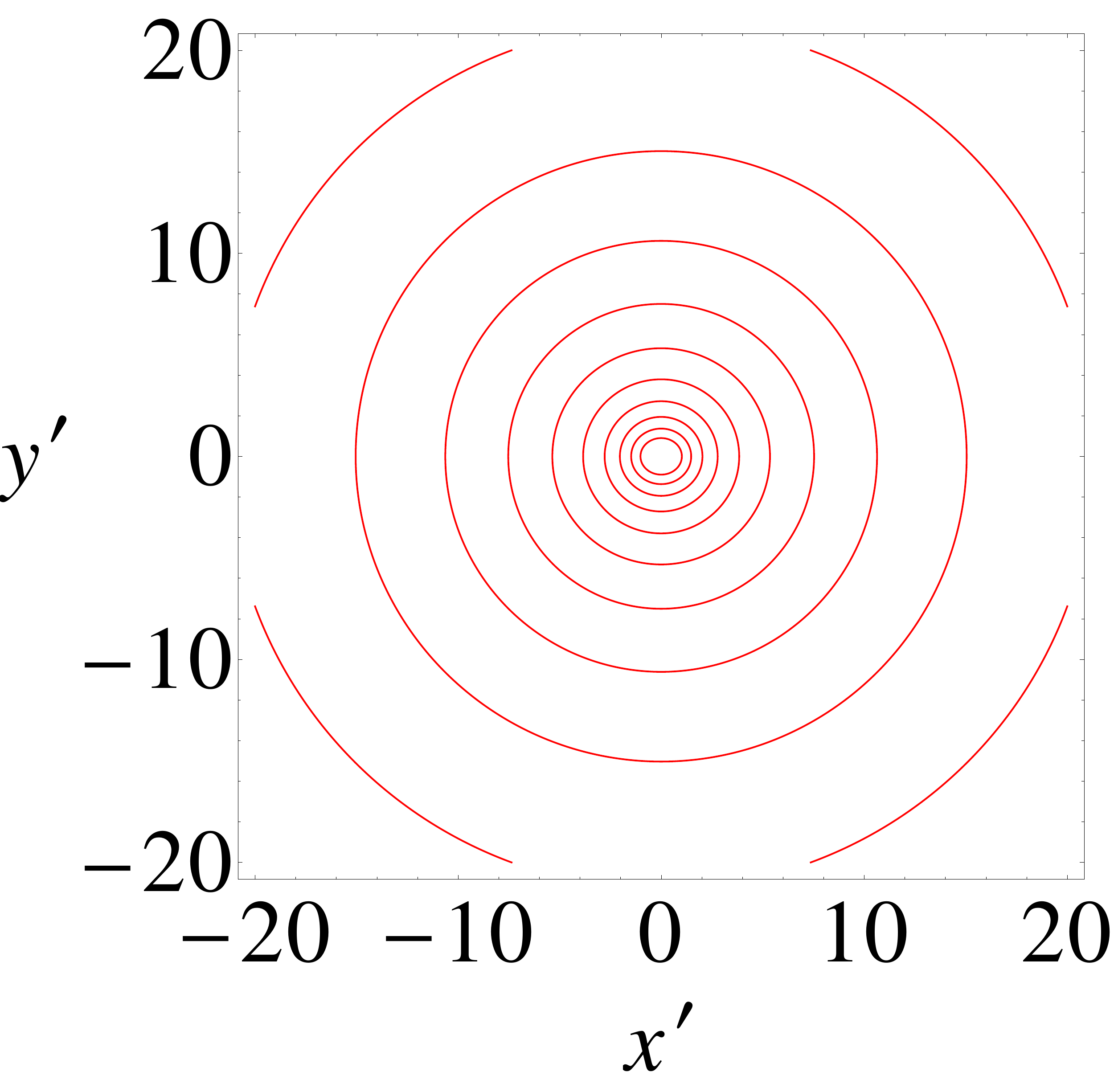}\quad\includegraphics[scale=0.1]{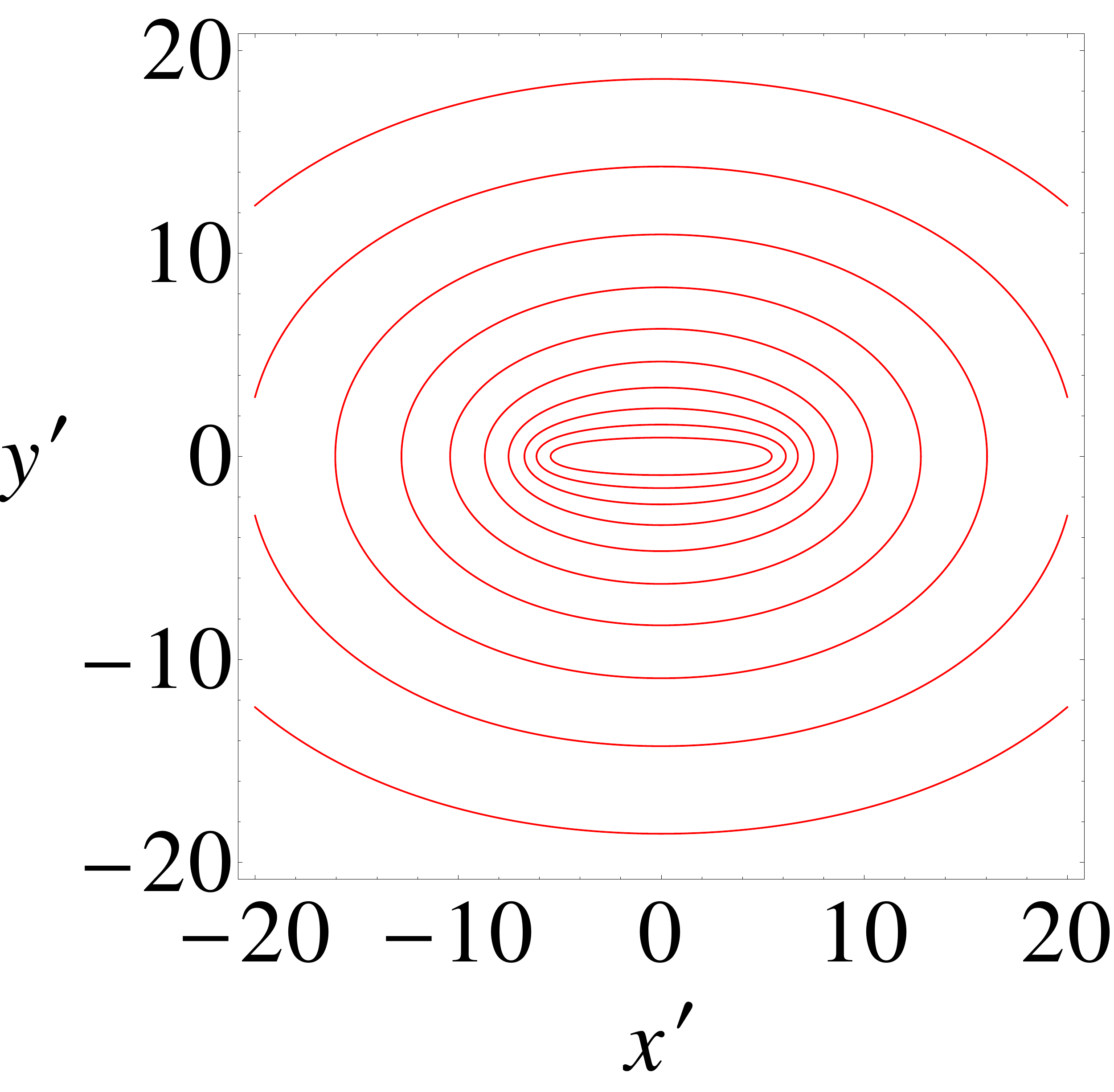}\quad\includegraphics[scale=0.1]{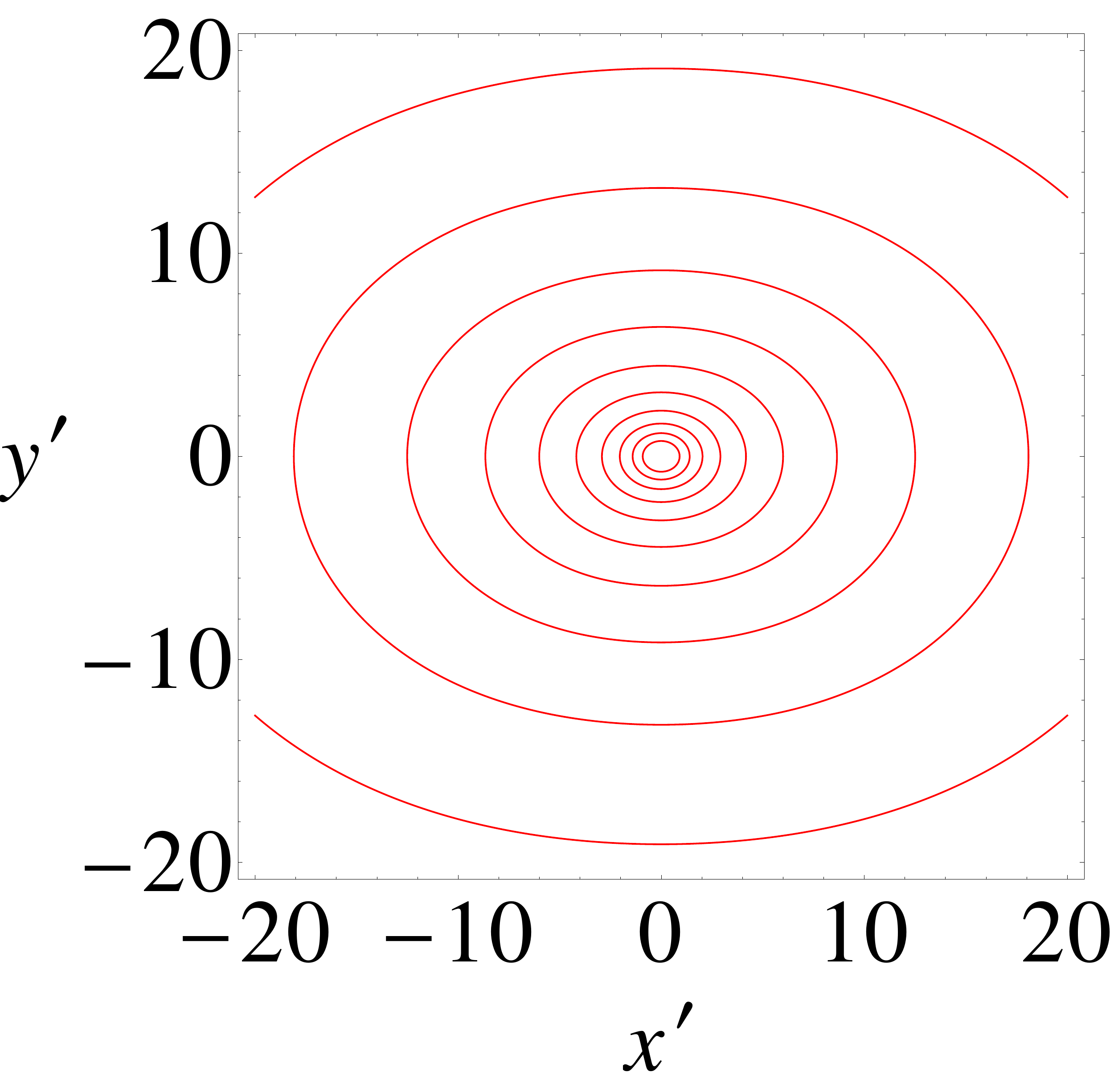}
  \caption{Logarithmically spaced contour plots of surface density for
    a model with $v_{0}=1$, $q=0.9$ and $a=6$. From left to right:
    $z'=z$ (downwards-on), $z'=y$ (edge-on), $z'=x$
    (face-on). Distances are given in units of $R_{\rm c}$.}
  \par\end{centering}
\label{fig:surface}
\end{figure}

\section{The Logarithmic Bar Model}

\subsection{Potential-Density Pair}

\citet{Lo92} introduced the idea of weighting spherical or
axisymmetric potentials with a needle-like density to give triaxial
potentials. They explored barred models based on underlying Plummer
(1911) or Miyamoto \& Nagai (1975) potentials. Despite succeeding work
by \citet{Vo10}, the technique has remained largely unexplored.  Here,
we let the underlying potential be denoted by $\Phi_{\rm L}$ and be of
logarithmic form
\begin{equation}
\Phi_{\rm
  L}(x,y,z)=\frac{v_0^{2}}{2}\ln(x^{2}+y^{2}+ z^{2}q^{-2} +R_{\rm
  c}^{2}),
\end{equation}
where $q$ is an axis ratio. This axisymmetric model has an
asymptotically flat circular velocity curve of amplitude $v_0$.  The
properties of this model are discussed in \citet{Ev93} and \citet{BT}.

The underlying potential $\Phi_{\rm L}$ is convolved with a thin
``needle'' density, $\lambda (x)$:
\begin{equation}
\Phi=\intop_{-\infty}^{\infty}\lambda(x')\Phi_{\rm L}(x-x',y,z)dx'.
\end{equation}
In this case, we choose $\lambda(x)$ to be a box function, spanning
$[-a,a]$ and normalised such that $\intop_{-a}^{a}\lambda(x)dx=1$. The
result is:
\begin{equation}
\Phi=\frac{v_0^{2}}{4a}\bigg[\chi\ln(\chi^{2}+T^{2})+2T\arctan\bigg(\frac{\chi}{T}\bigg)\bigg]_{x-a}^{x+a},
\end{equation}
where $T=\sqrt{R_{\rm c}^{2}+y^{2}+z^{2}/q^{2}}$, and we have adopted
the notation:
\begin{equation}
\bigg[f(\chi)\bigg]_{x_{1}}^{x_{2}}\equiv f(x_{2})-f(x_{1}),
\end{equation}
which we shall use throughout the paper. This arises naturally because
the operation used to obtain the potential is a definite integral. 
The corresponding density can then be obtained using Poisson's equation:
\begin{eqnarray}
\rho&=&\frac{v_0^{2}}{8\pi
  Gaq^{2}T^{2}}\bigg[\frac{(q^{2}R_{c}^{2}+(1-q^{-2})z^{2})\chi}{\chi^{2}+T^{2}}\nonumber\\ &
  &\qquad\qquad
  +\frac{\big((1+q^{2})R_{c}^{2}+y^{2}+z^{2}\big)\arctan\bigg(\dfrac{\chi}{T}\bigg)}{T}\bigg]_{x-a}^{x+a}.
\label{eq:bardens}
\end{eqnarray}
We shall refer to this model as the {\it logarithmic bar}. The density
represents the total self-gravitating matter, both luminous and dark.
It is highly triaxial near the centre, but becomes axisymmetric at
large radii. A necessary and sufficient condition for positive density
is $q>1/\sqrt{2}$, just as for the logarithmic potential itself (see
\citealt{BT}).  Furthermore, if $R_{\rm c}=0$, then the density
diverges along the $x$-axis, which is undesirable. For practical
purposes, therefore, $R_{\rm c} >0$.  Sample density contours are
shown in Fig.~\ref{fig:dens}, from which we can see that the model can
attain highly triaxial and extended shapes. Within the bar regime, the
density on the major axis falls typically by a factor of $\sim 2$ from
the centre to the end of the bar ($|x| \sim a$).  At large distances,
the density falls off like
\begin{equation}
  \rho(x,0,0) \sim {v_0^2\over 4 \pi G q^2 x^2},\qquad
  \rho(0,0,z,) \sim {v_0^2 (2- q^{-2})\over 4 \pi G z^2}.
\end{equation}
We see that, far from the bar, the model looks like the axisymmetric
logarithmic potential. To represent a barred galaxy, we keep in mind
the picture that the inner parts are luminous, whilst the outer parts
are dominated by dark matter which provides the flat rotation curve.

The gravitational forces are:
\begin{align}
F_{x}=&\quad-\bigg[\frac{v_{0}^{2}}{4a}\ln(\chi^{2}+T^{2})\bigg]_{x-a}^{x+a},
\nonumber\\
F_{y}=&\quad-\bigg[\frac{v_{0}^{2}y}{2aT}\arctan\bigg(\frac{\chi}{T}\bigg)\bigg]_{x-a}^{x+a},\\
F_{z}=&\quad-\bigg[\frac{v_{0}^{2}z}{2aq^{2}T}\arctan\bigg(\frac{\chi}{T}\bigg)\bigg]_{x-a}^{x+a}.\nonumber
\end{align}
We have written this out explicitly to emphasise the fact that orbit
integration in the model is very fast, as the force components contain
only elementary functions. 

The model contains four parameters: $v_{0}$, $a$, $R_{c}$ and $q$. The
physical interpretations of these constants are as follows: $v_{0}$
remains the asymptotic rotation speed in the galactic plane, just as
for the original logarithmic potential; $q$ is the axis ratio of the
equipotentials in the $(y,z)$ plane (as well as the axis ratio at
large distances from the origin in the $(x,z)$ plane); $a$ has the
simple interpretation of being the half-length of the bar. Finally,
the ratio $a/R_{c}$ is related to the axis ratios of the bar in the
$(x,y)$ and $(x,z)$ planes, in the sense that a larger value of
$a/R_{\rm c}$ corresponds to more extreme axis ratios. This degree of
freedom may be removed, however, by working with $R_{\rm c}$ as the
unit length: this will be the case for the remainder of this paper. In
these units, $a$ both quantifies the half-length of the bar and the
aforementioned axis ratios.

We remark that our four-parameter triaxial bar model compares very
favourably with the competition! A triaxial Ferrers bar already has
four free parameters, and it still has to be embedded in an
axisymmetric background which normally has at least a further two or
three~\citep{Te85, Ka05}. The bar models devised by \citet{Lo92} also
needed to be immersed in a background model for practical
use~\citep{Ha16}.

\subsection{Equipotentials}

Unlike the case of a Ferrers bar, the equipotentials of this model do
not lie on similar concentric ellipsoids. Yet, an axis ratio is still
a useful quantification of the geometry of the bar. If we Taylor
expand the potential around the origin, we find:
\begin{equation}
\Phi=\frac{v_{0}^{2}}{2}\bigg(\frac{x^{2}}{a^{2}+1}+\frac{(y^{2}+z^{2}/q^{2})\arctan[a]}{a}\bigg)+\ldots
\label{eq:taylor}
\end{equation}
We can clearly see that, close to the centre, the equipotentials
approximately lie on concentric ellipsoids, and hence meaningful axis
ratios may be extracted. The $(y,z)$ axis ratio is simply $q$, but the
$(x,y)$ axis ratio is less trivial.

A prescription for choosing a value of $a$ that approximately satisfies
a given axis ratio in the $(x,y)$ plane, $p_{\Phi}$, is readily given as
\begin{eqnarray}
a&=&\tan\theta, \nonumber\\
\theta &=&\bigg[\frac{7}{2}-\frac{7^{2/3}3}{2\zeta}+\frac{7^{1/3}}{2}\zeta\bigg]^{1/2}\\
\zeta &=& \bigg[-11-45p_{\Phi}^{2}+\sqrt{5}\sqrt{62+198p_{\Phi}^{2}+405p_{\Phi}^{4}}\bigg]^{1/3}\nonumber
\end{eqnarray}
The axis ratio is exactly $\sqrt{\mathrm{sinc}\,2\theta}$. This
prescription is very effective, overestimating $p_{\Phi}$ by less than
one per cent down to an axis ratio of $p_{\Phi}\simeq 0.5$, rising to
approximately six per cent when $p_{\Phi}\simeq 0.3$.

The approach of Taylor expanding close to the origin is not effective
in the case of the density, both because the expansion is more
complicated than that of the potential, and because the density is
more flattened than the potential in this region. Since the axis
ratios of the density are of interest because they represent the
physical dimensions of the bar itself, we suggest the following
convention: we use the values on the $y$ and $z$ axes at which the
density is equal to $\rho(a,0,0)$ to compute the axis ratios, since
this is consistent with the physical interpretation of $a$ as the
half-length of the bar. We note that this must be done numerically,
and that the behaviour scales roughly as $p_{\rho}\sim(1+a)^{-1}$.

\begin{table}
\begin{tabular}{lcccr}
\hline
Model & $p_{\Phi}$ & $p_{\rho}$ & $a$ & $\Omega_{b}$ \tabularnewline
\hline
Weak Bar & 0.5 & 0.46 & 2.85 & 0.33\tabularnewline
Strong Bar & 0.3 & 0.14 & 7.55 & 0.13\tabularnewline
\hline
\end{tabular}\caption{Parameters for the two models used. In both cases $v_{0}=1$.}
\label{table:models}
\end{table}

\begin{figure}
\includegraphics[scale=0.2]{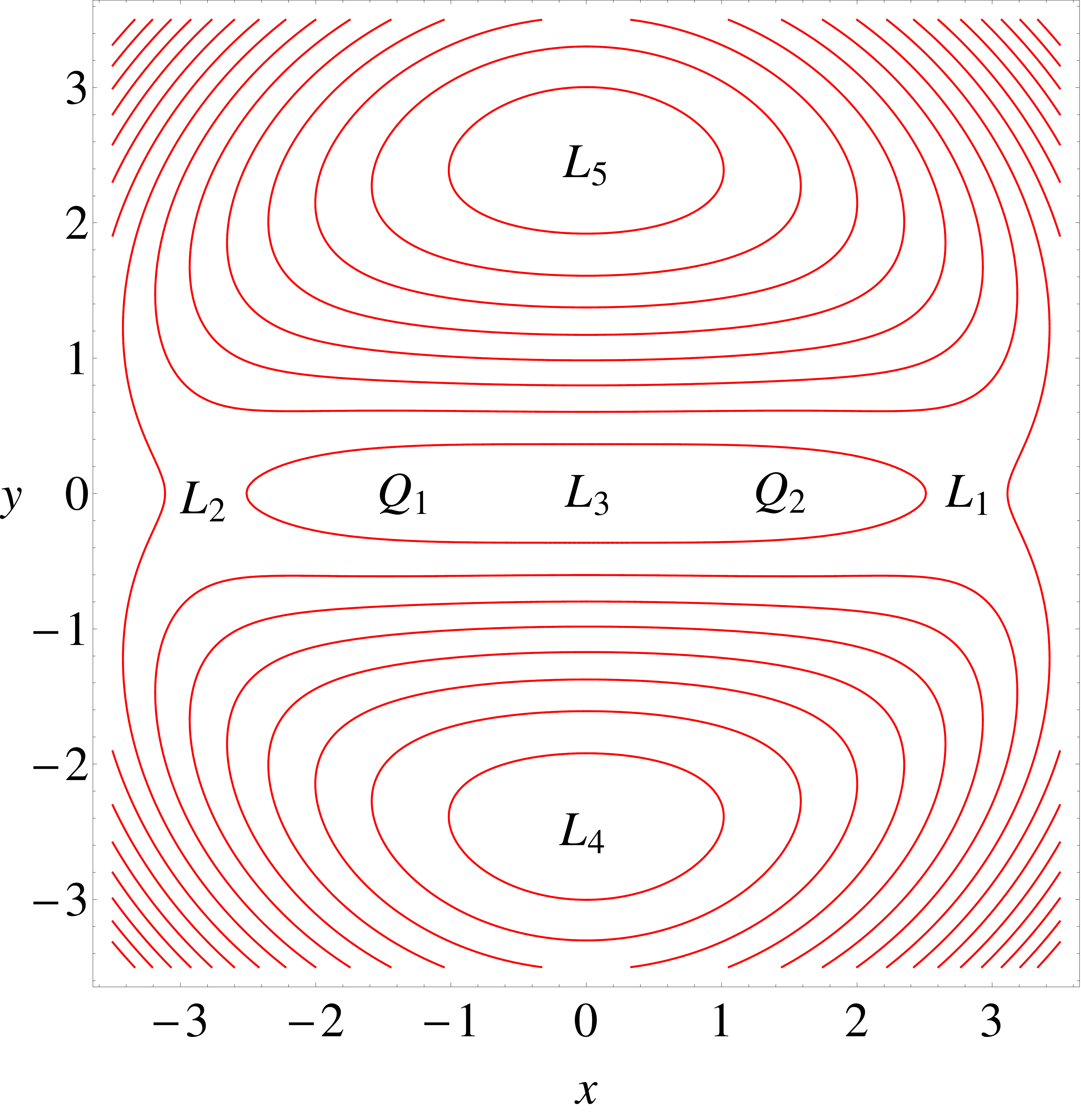}\caption{Contours of the
  effective potential for the weak bar model. $L_{4}$ and $L_{5}$
  are maxima, $L_{1}$ and $L_{2}$ are saddle points, and $L_{3}$ is a
  minimum. If $\Omega_{b}>\Omega_{\rm crit}$, $L_{3}$ is a saddle
  point and the minima $Q_{1}$ and $Q_{2}$ appear.}
\label{fig:phieff}
\end{figure}  

\begin{figure*}
\begin{centering}
\includegraphics[scale=0.11]{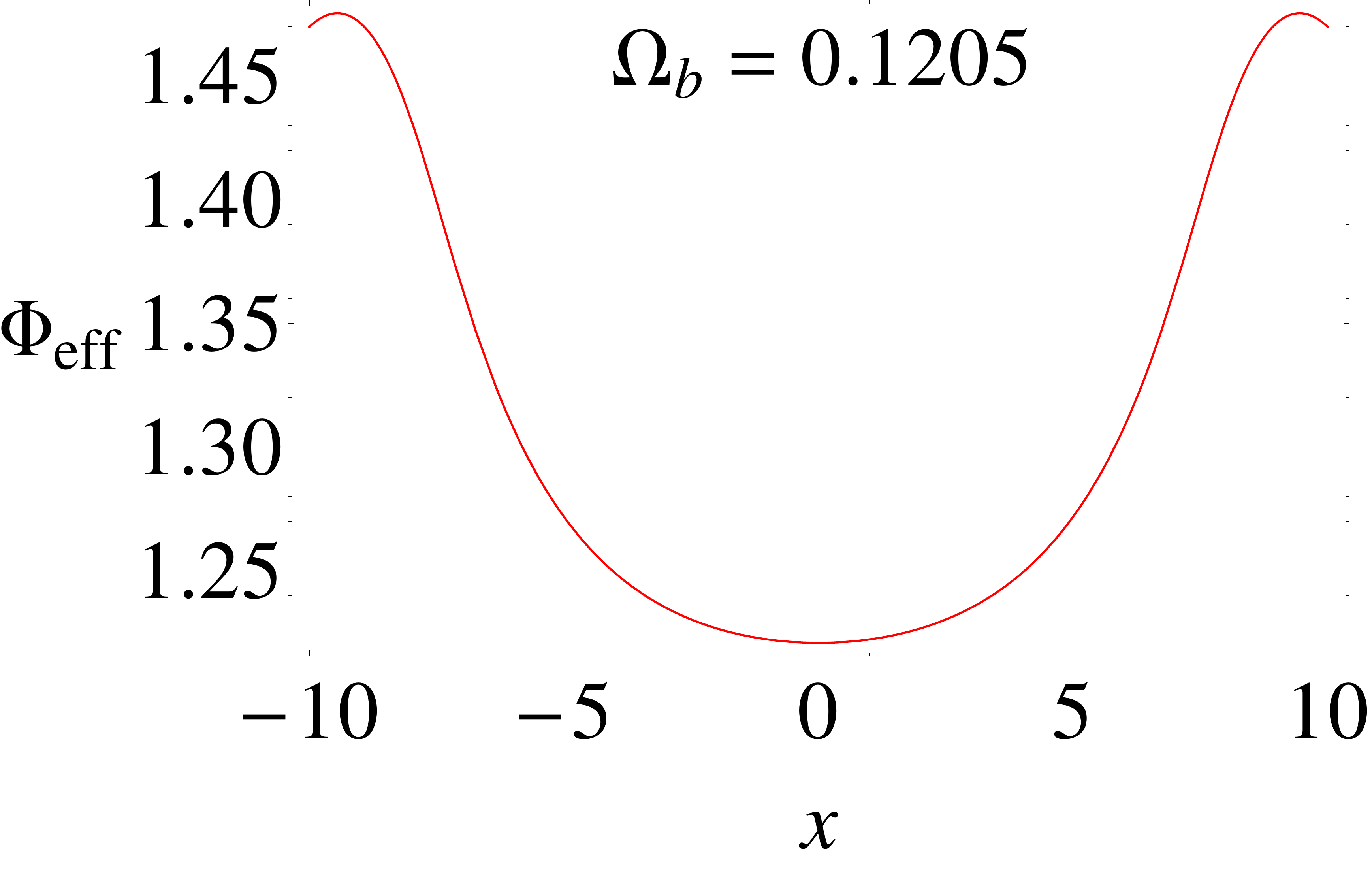}\quad\includegraphics[scale=0.11]{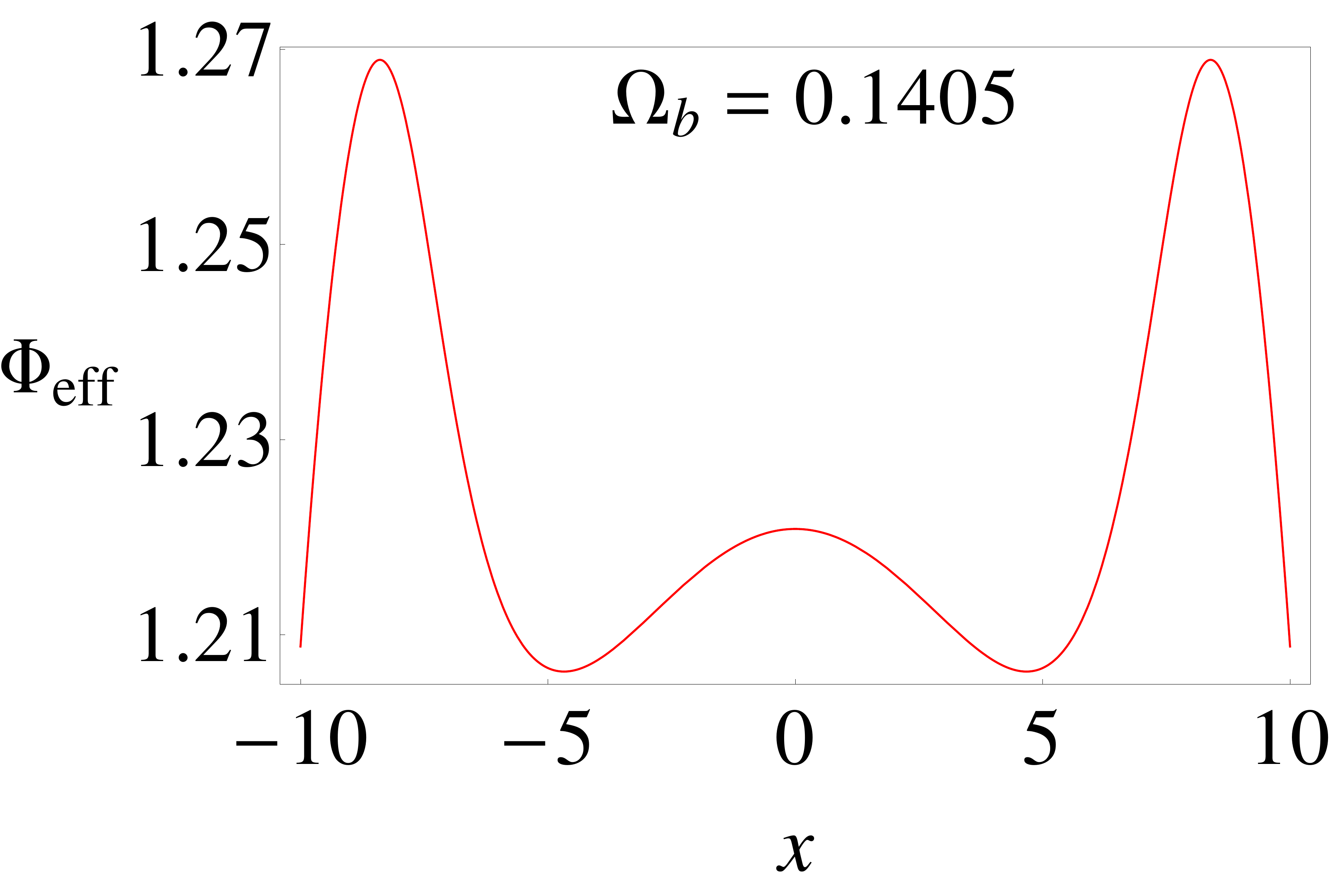}\quad\includegraphics[scale=0.11]{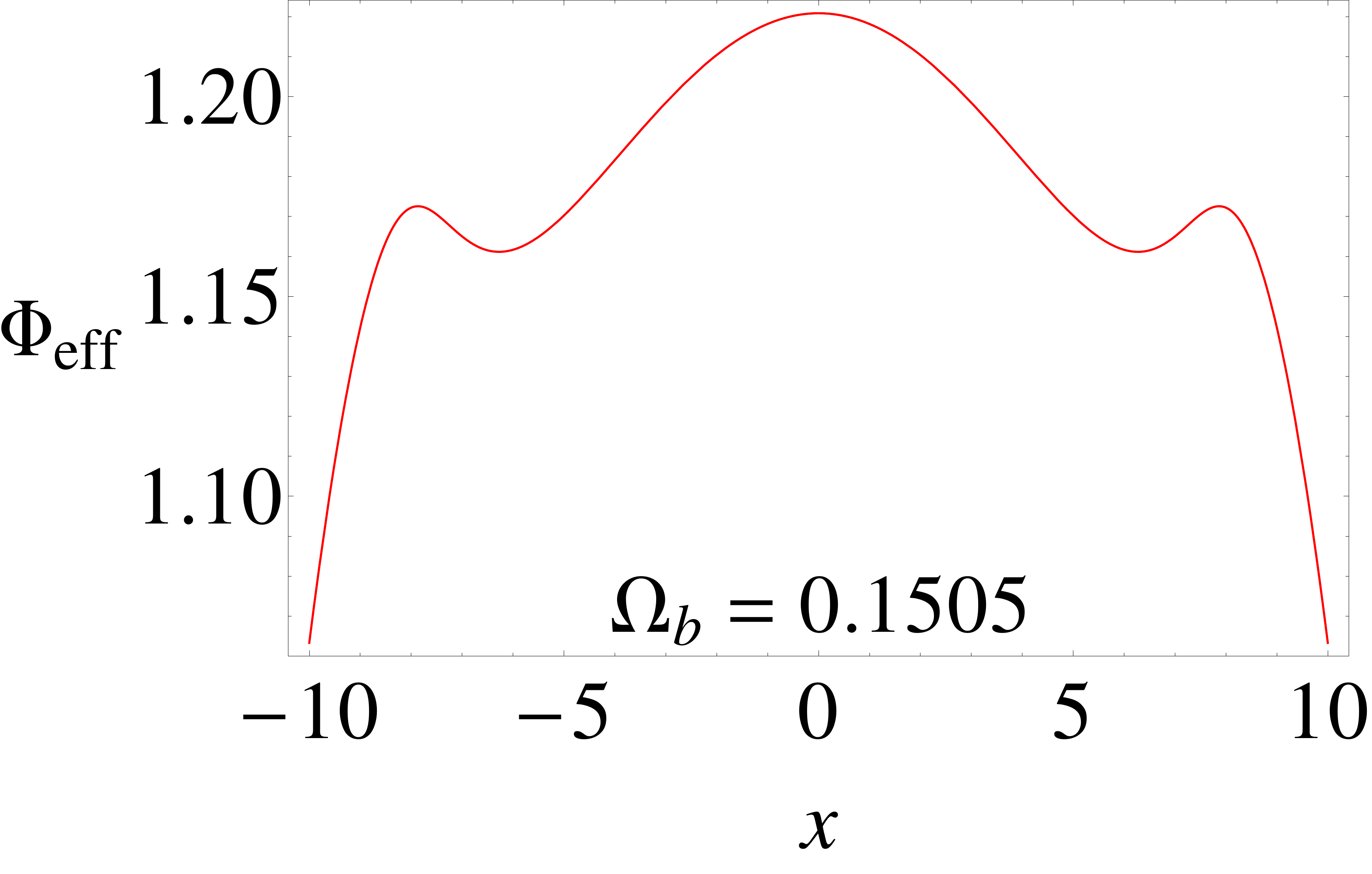}
\quad\includegraphics[scale=0.11]{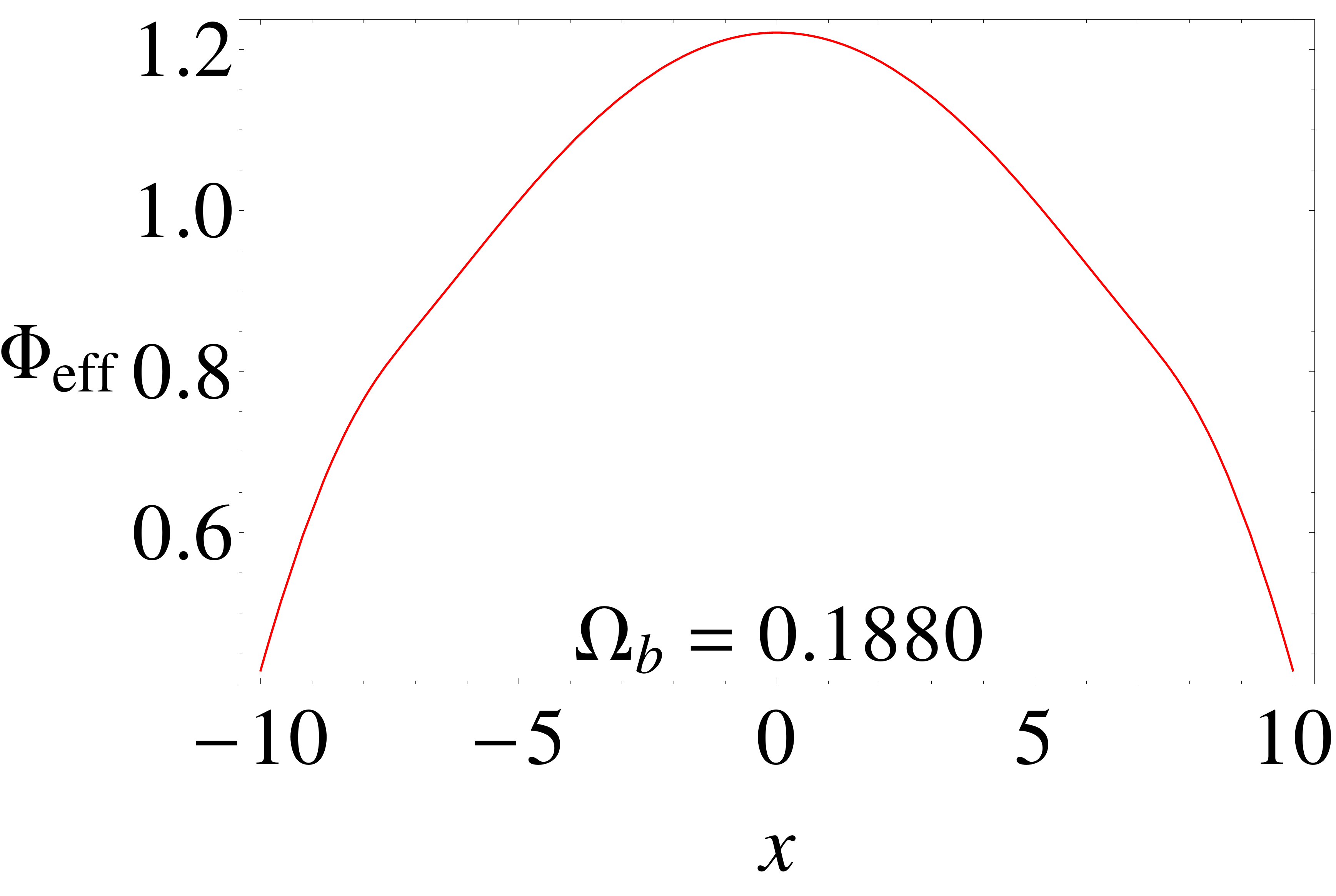}\caption{The
  evolution of the effective potential along the $x$-axis with
  $\Omega_{b}$ in the strong bar model, for which $\Omega_{\rm
    crit}=0.1313$. Notice the emergence of additional stationary
  points in the third panel.}
\label{fig:effpotx}
\par\end{centering}
\end{figure*}

\subsection{Surface Density}

To evaluate the surface density along a general line of sight, we must
transform from internal coordinates ($x,y,z$) to projected coordinates
($x',y',z'$). For a triaxial system, there are two viewing angles
$\vartheta$ and $\varphi$. The required transformation is~\citep[see
  e.g.,][]{Ev00}:
\begin{equation}
\left(\begin{array}{c}
x\\
y\\
z\end{array}\right)=\left(\begin{array}{ccc}
-\sin\vartheta & -\cos\varphi\sin\vartheta & \cos\varphi\sin\vartheta\\
\cos\varphi & -\sin\varphi\cos\vartheta & \sin\varphi\sin\vartheta\\
0 & \sin\vartheta & \cos\vartheta\end{array}\right)\left(\begin{array}{c}
x'\\
y'\\
z'\end{array}\right).
\end{equation}
For arbitrary viewing angles, the integration along the line of sight
needs to be done numerically. However, when the line of sight
coincides with one of the principal axes of the figure, the results
are analytic.

This illustrates an advantage of this method of creating barred
potentials. Any quantity that is computed using a linear operation on
$\Phi$ can be first computed for the kernel (in this case $\Phi_{\rm
  L}$), and then convolved with the needle density.  Hence, we can
compute the surface densities for the barred model using the results
from the axisymmetric model. We emphasise that the constraint $q>
1/\sqrt{2}$ is necessary to ensure that the model is physical and has
everywhere positive three dimensional density.  Projecting along the
$z'=z$ axis gives (in units $R_{\rm c}=1$):
\begin{equation}
\Sigma (x',y')=\frac{qv_0^{2}}{8aG(1+y'^{2})}\bigg[\frac{\chi}{P(\chi)}+(1+{y'}^{2})\ln[\chi+P(\chi)]\bigg]_{x'-a}^{x'+a},
\end{equation}
where $P(\chi)=\sqrt{1+\chi^{2}+y'^{2}}$. Here, the barred galaxy is
being viewed face-on.

Projecting along the $z'=y$ axis gives:
\begin{eqnarray}
\Sigma(x',y')=\frac{v_0^{2}}{8aGq^{2}(1+{y'}^{2}/q^{2})}\bigg[\frac{\chi(q^{2}+(1-q^{-2}){y'}^{2})}{Q(\chi)}\nonumber \\
+(1+{y'}^{2}/q^{2})\ln[\chi+Q(\chi)]\bigg]_{x'-a}^{x'+a},
\end{eqnarray}
where $Q(\chi)=\sqrt{1+\chi^{2}+{y'}^{2}/q^{2}}$. This expression is
useful when considering edge-on galaxies, which are a topic of much
interest observationally. The surface density profile along the bar
major axis is flattish. Typically, the surface density falls from its
central value by a factor of $\lesssim 2$ by the end of the bar along
the major axis ($|x'|\sim a$).  Along the minor axis at $|y'|\sim a$,
the fall-off factor is of course much steeper ($\sim 5$, depending on
the choices of $q$ and $a$).

The final projection along a principal axis is $z' = x$, which yields:
\begin{equation}
\Sigma(x',y')=\frac{v_0^{2}}{4Gq^{2}T^{3}}\Big(1+q^{2}+{x'}^{2}+{y'}^{2}\Big).
\end{equation}
Plots of the surface density are shown in Fig. 2 for the face-on,
edge-on and downwards-on cases. Of course, this is the total (luminous
and dark) surface density.

\section{Orbital Structure}

Here, we provide an analysis of the in-plane orbital structure (with
figure rotation) of this potential. The planar periodic orbits that
exist at various energies play a crucial role. They sire orbital
families that librate around the periodic orbits. Ultimately, this
information gives us insight into how the stars moving along these
different orbits generate the density distribution, and hence the
potential.

The Hamiltonian is:
\begin{eqnarray}
H_{\rm J}&=&\frac{1}{2}(\dot{x}^2+\dot{y}^2)+\Phi_{\rm eff}(x,y),\\
\Phi_{\rm
  eff}(x,y)&=&\Phi(x,y)-\frac{1}{2}\Omega_{b}^2(x^2+y^2),\nonumber
\label{eq:ham}
\end{eqnarray}
where $\Omega _{\rm b}$ is the pattern speed and the phase space
coordinates are written in terms of the frame in which the potential
is static. The quantity $\Phi_{\rm eff}$ is known as the effective
potential. The Hamiltonian is conserved in the rotating frame and is
called the Jacobi constant (which we shall loosely refer to as the
energy). The equations of motion are then extracted using Hamilton's
equations, and orbits are integrated numerically. 

Note that the potential of the whole model is steady in a frame
rotating with pattern speed $\Omega_{\rm b}$. So, the figure of the
inner triaxial bar rotates. However, the outer parts are axisymmetric,
and there is no difference between figure rotation and orbital
streaming in axisymmetry. So, the orbital streaming (or the odd part
of the distribution function) can be chosen so that there is no net
rotation of the outer parts.

\begin{figure}
\includegraphics[width=\columnwidth]{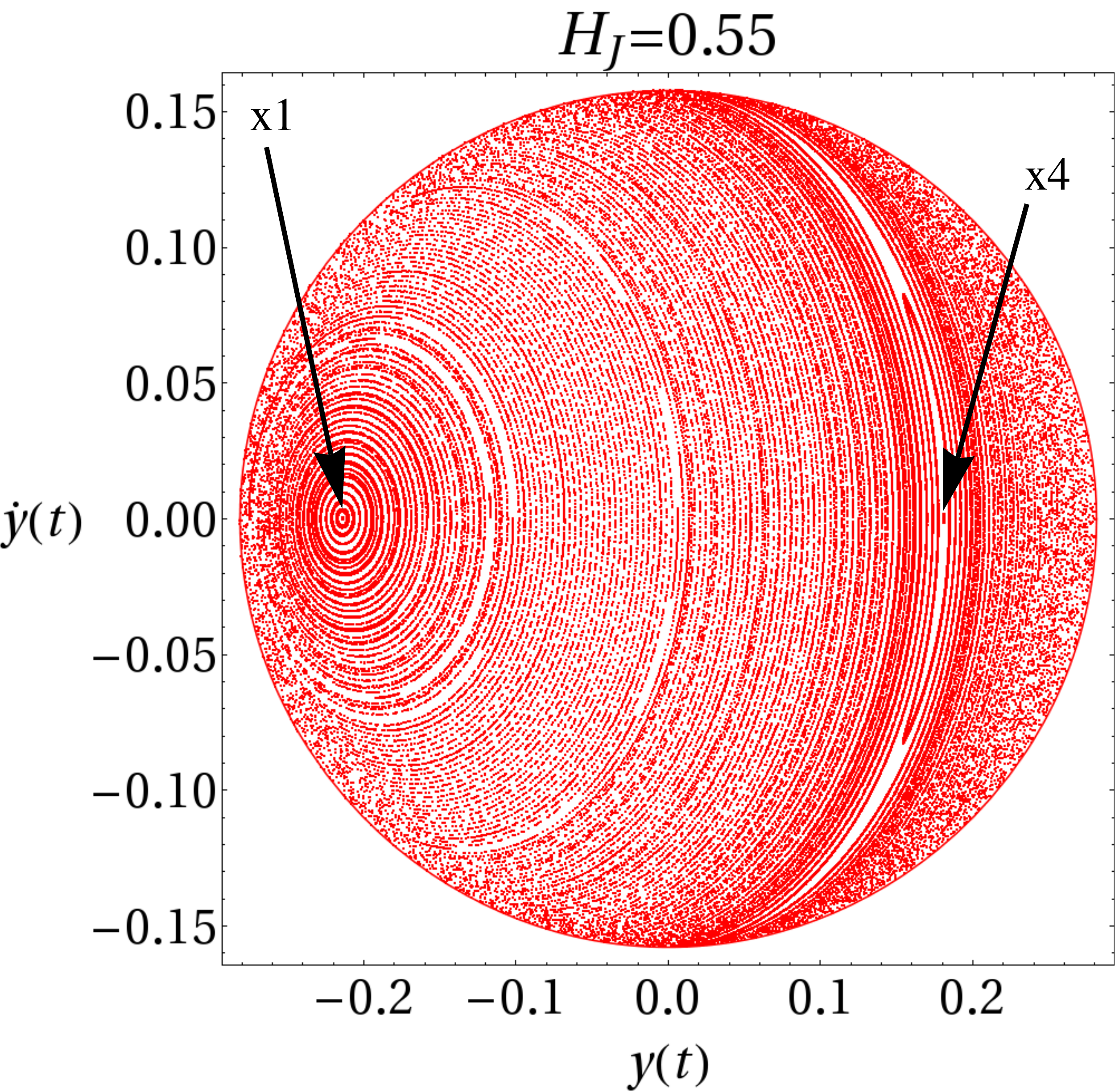}
\caption{A surface of section for the weak bar model at Jacobi energy
  $H_J = 0.55$. The location of the $x_1$ and $x_4$ periodic orbits are
  marked. }
\label{fig:sos_weak}
\end{figure}

\begin{figure}
\includegraphics[width=\columnwidth]{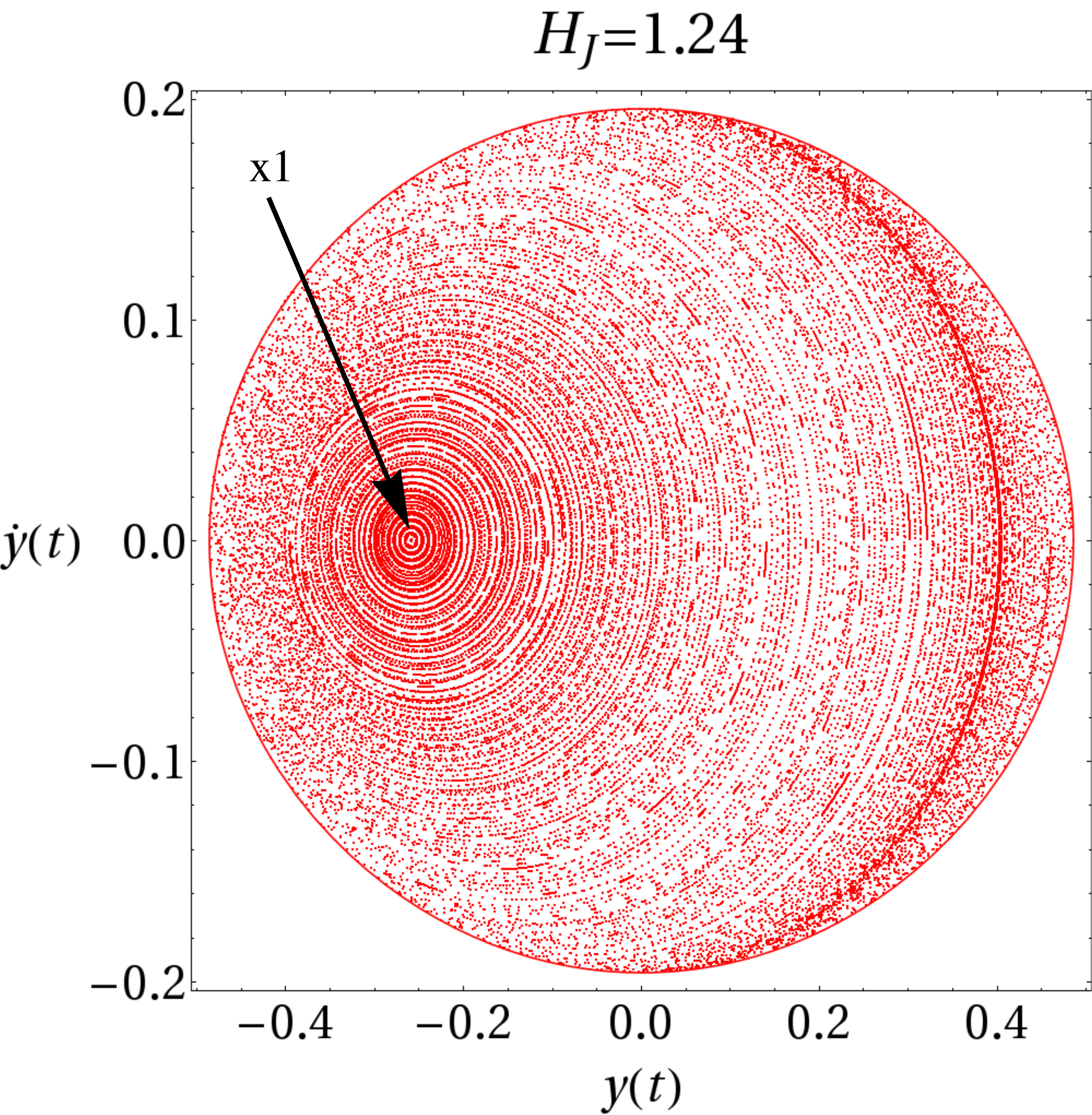}
\caption{A typical surface of section for the strong bar model. At
  this energy ($H_J = 1.24$), the $x_1$ periodic orbit and its family
  dominates. There are no stable $x_4$ orbits.}
\label{fig:sos_strong}
\end{figure}

\begin{figure}
\includegraphics[width=\columnwidth]{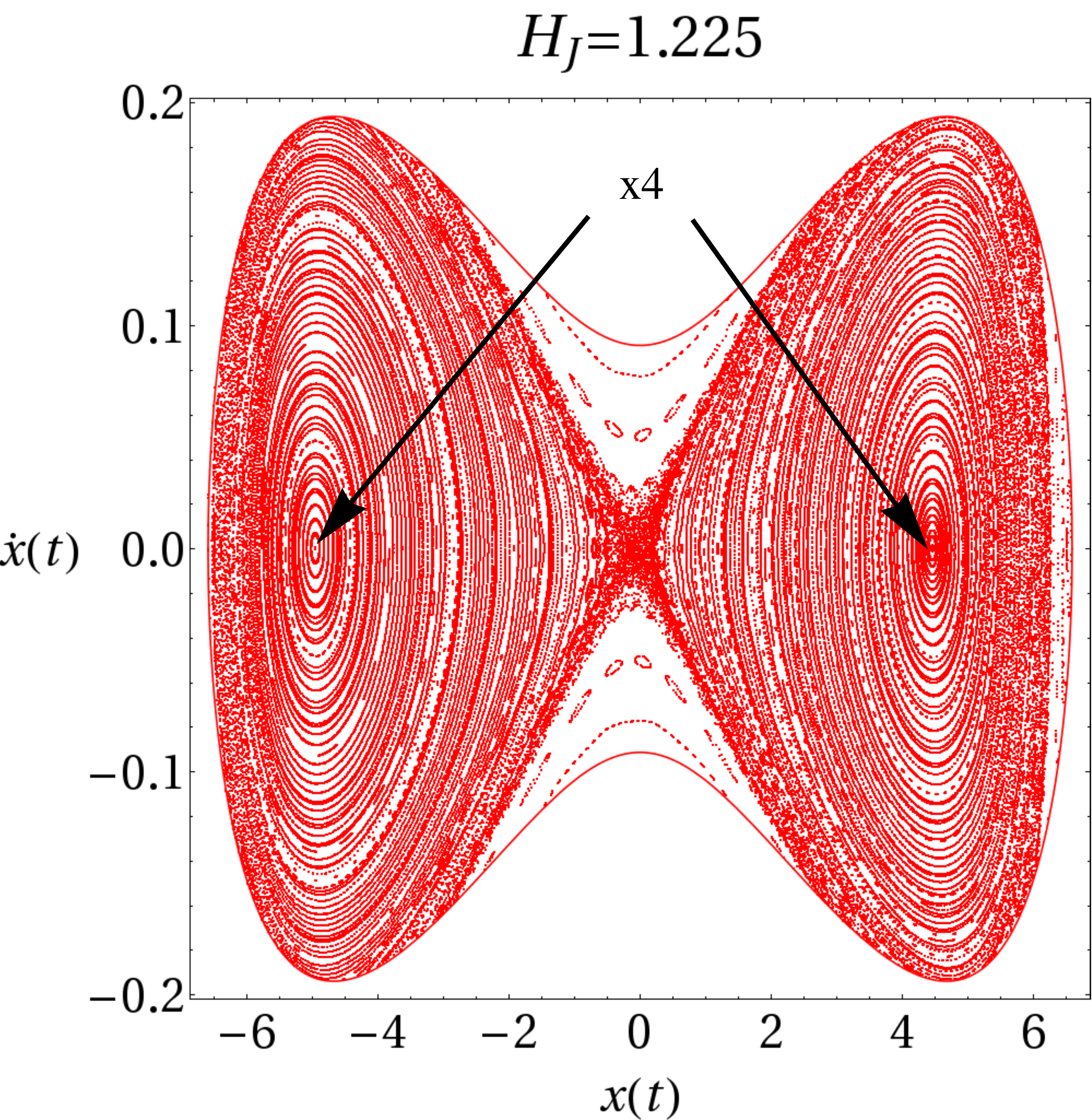}
\caption{A surface of section of the strong bar model, but with
  $\Omega_{b}>\Omega_{\rm crit}$ ($\Omega_{b}=0.1405$), in the regime
  when $Q_{1}$ and $Q_{2}$ exist. The periodic orbits are two
  off-centered $x_4$ orbits that circulate around $Q_{1}$ and $Q_{2}$.
  There is a chaotic layer associated with the separatrix.}
\label{fig:sos_extra}
\end{figure}

\begin{figure*}
\includegraphics[scale=0.25]{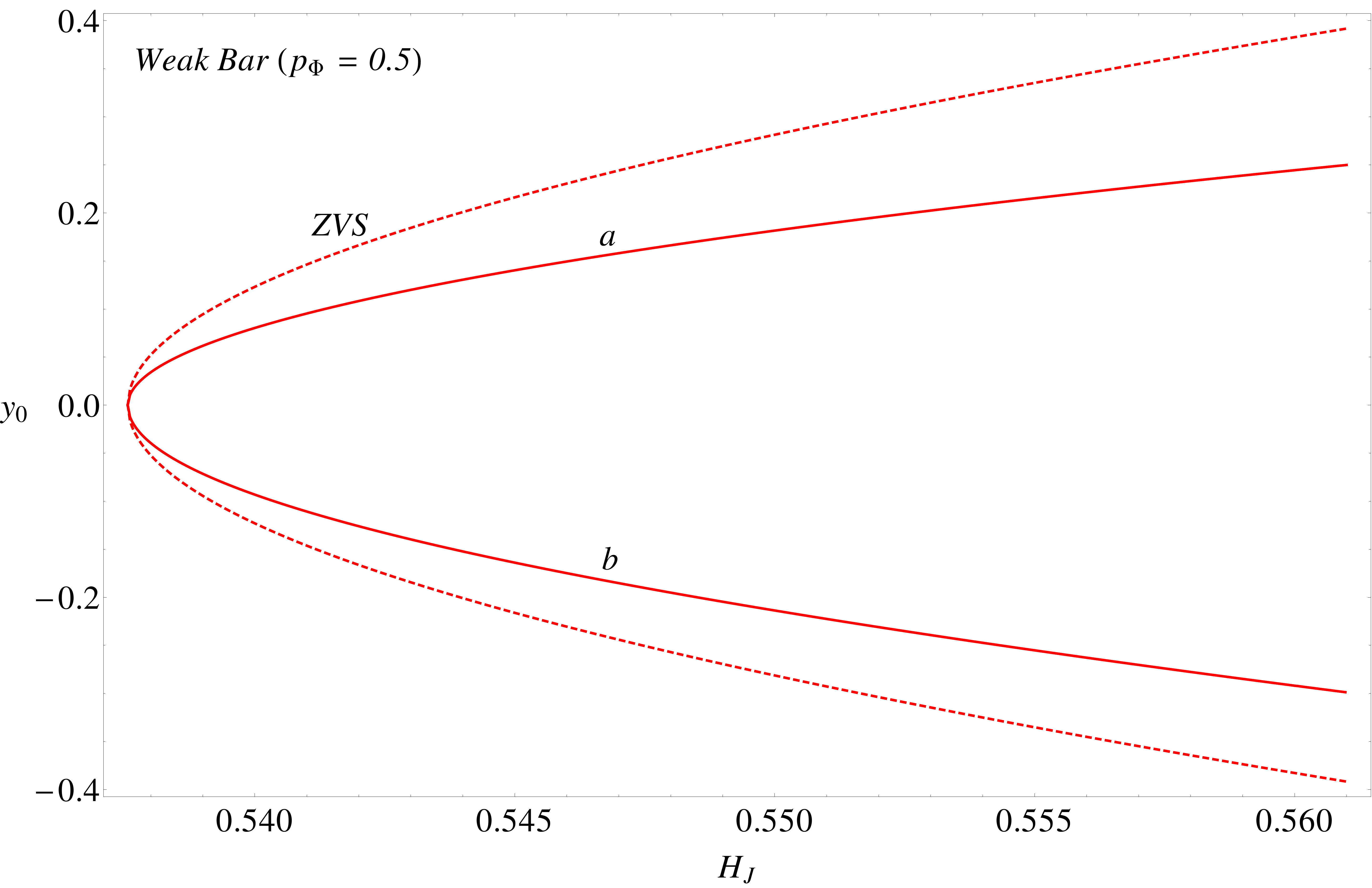} \\
\begin{centering}
\includegraphics[scale=0.2]{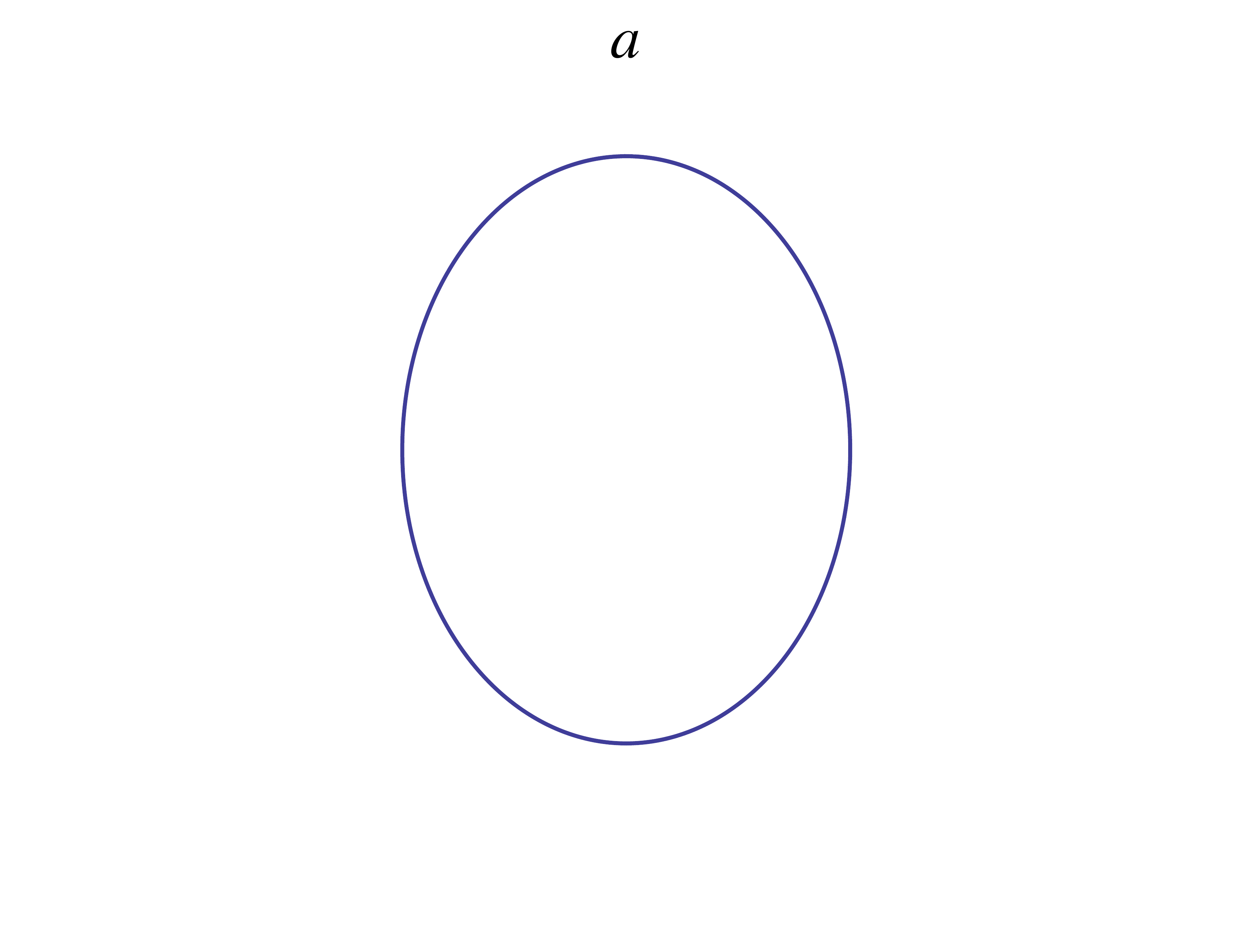}\hspace{6em}\includegraphics[scale=0.2]{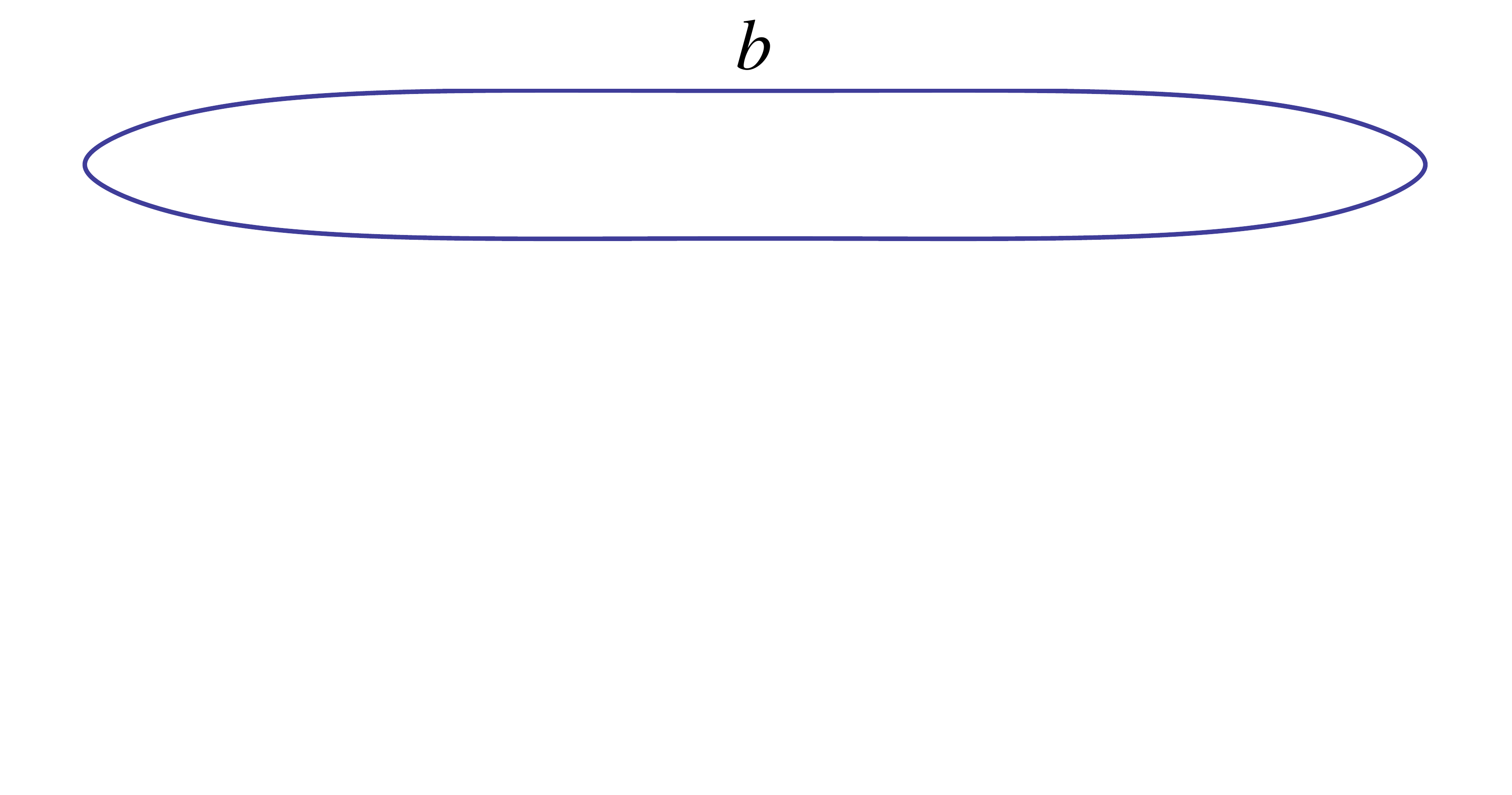}\\
\caption{Characteristic diagram of stable orbits with multiplicity 1
  for the weak bar model. This potential has a very simple orbital
  structure, with the retrograde $x_4$ ($a$) and prograde $x_1$ ($b$) orbits
  dominating.}
\label{fig:charweak}
\par\end{centering}
\end{figure*}

\begin{figure*}
\includegraphics[scale=0.25]{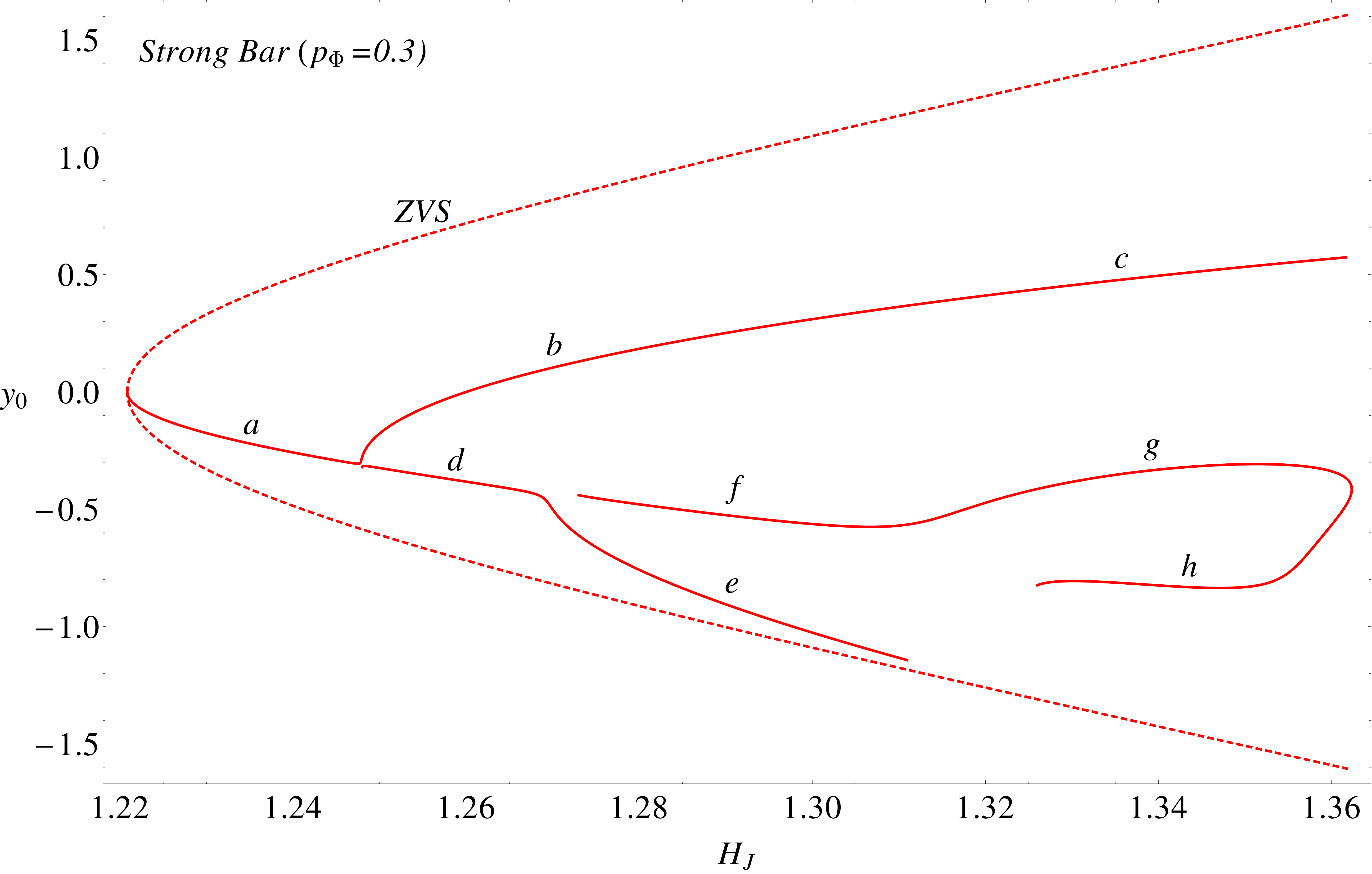} \\
\begin{centering}
\vspace{1pt}\includegraphics[scale=0.2]{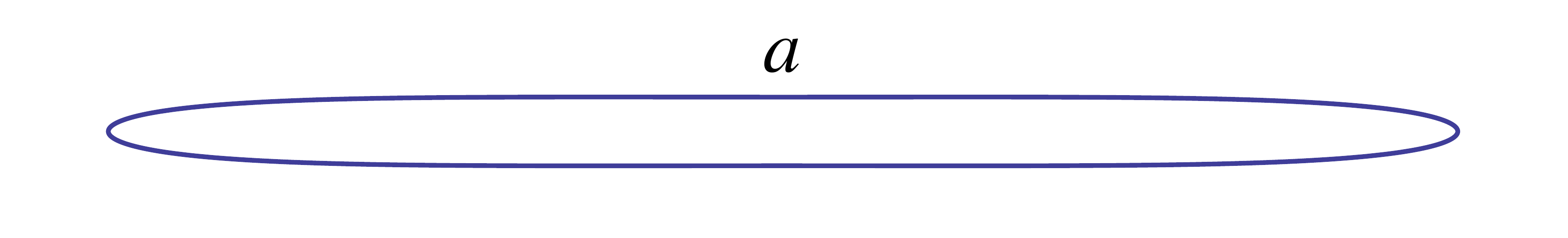}\hspace{6em}\includegraphics[scale=0.2]{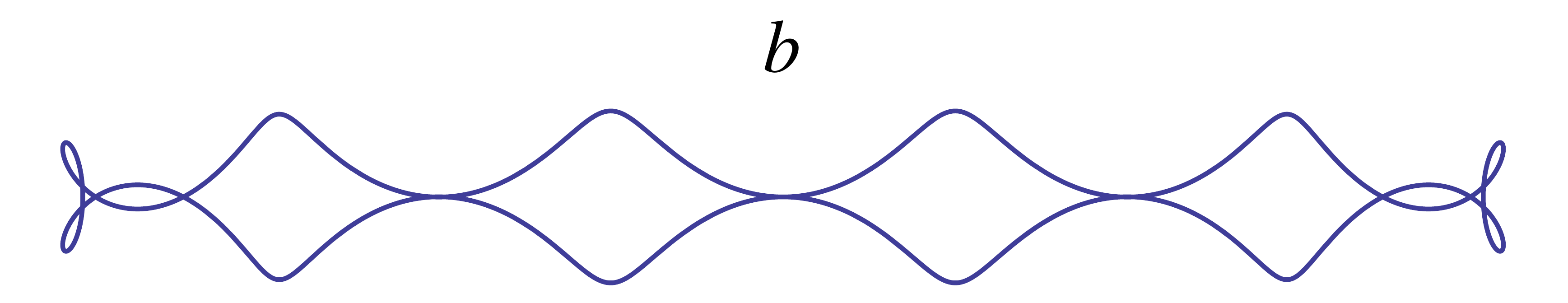}\\
\includegraphics[scale=0.2]{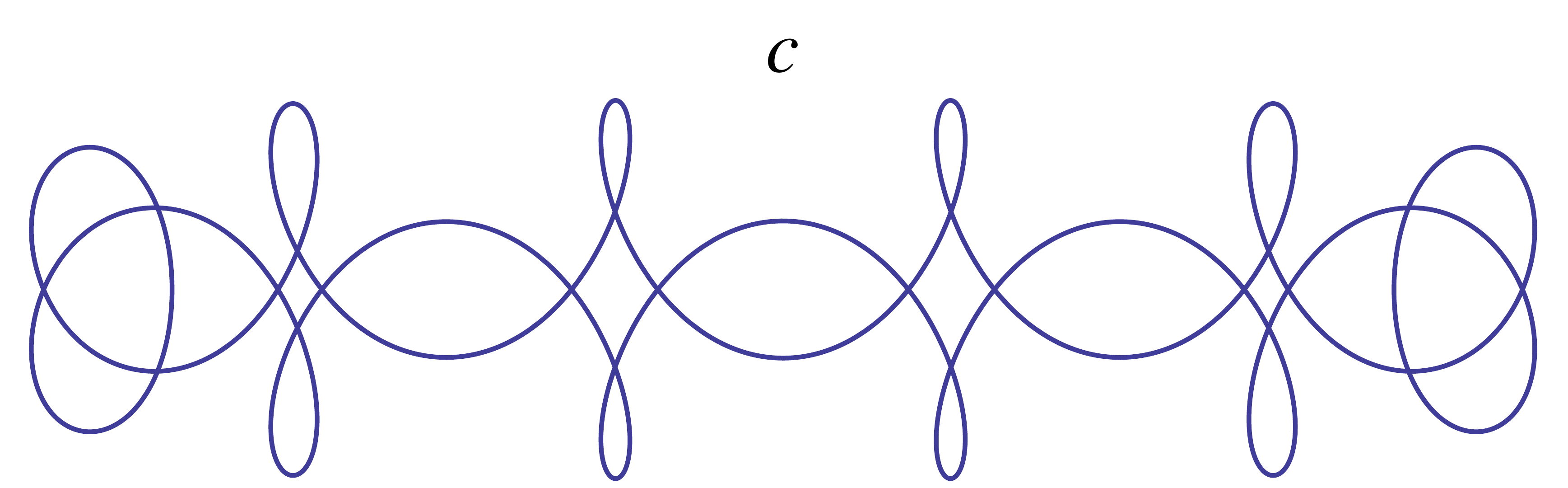}\hspace{6em}\includegraphics[scale=0.2]{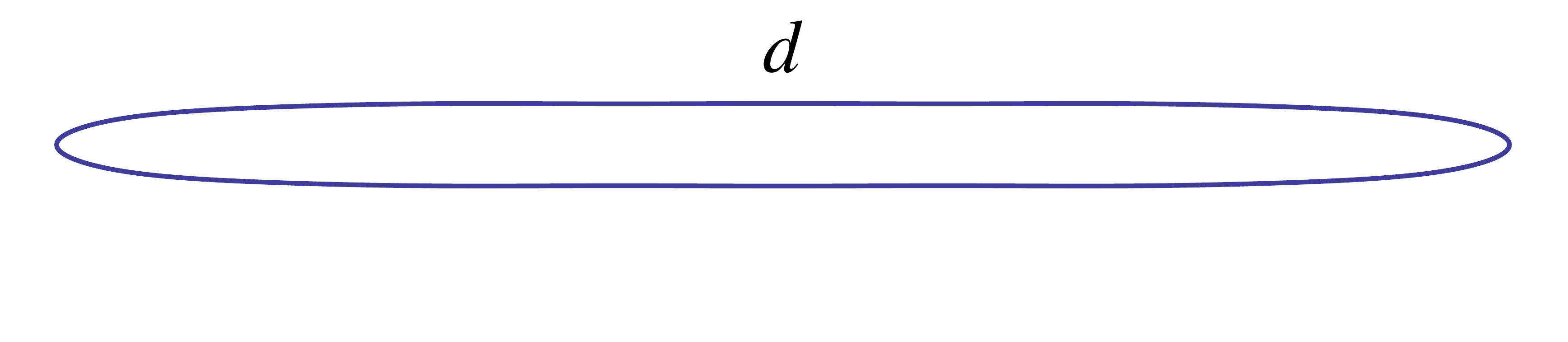}\\
\includegraphics[scale=0.2]{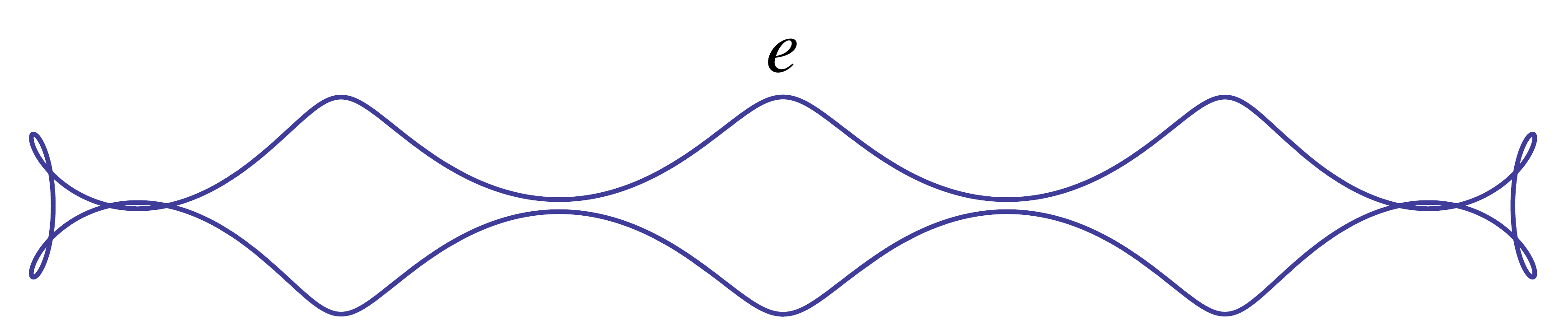}\hspace{6em}\includegraphics[scale=0.2]{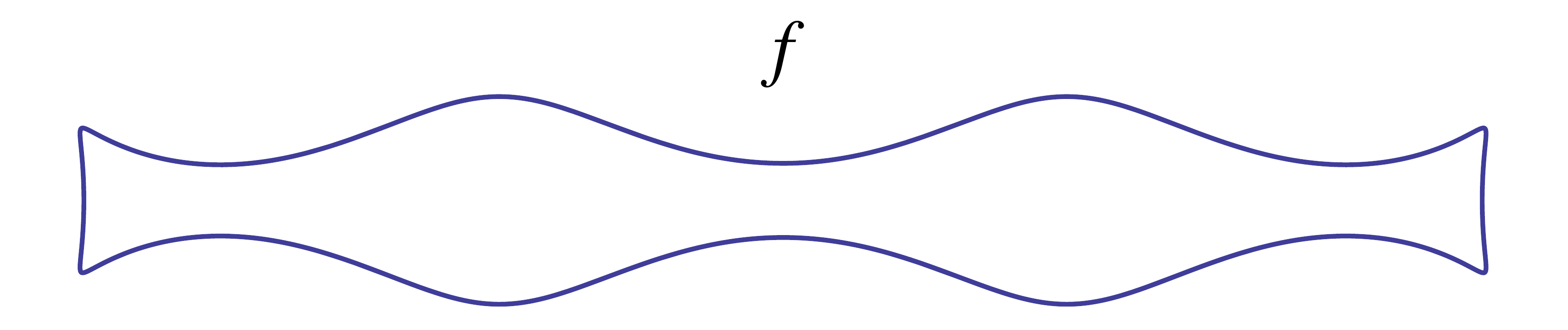}\\
\includegraphics[scale=0.2]{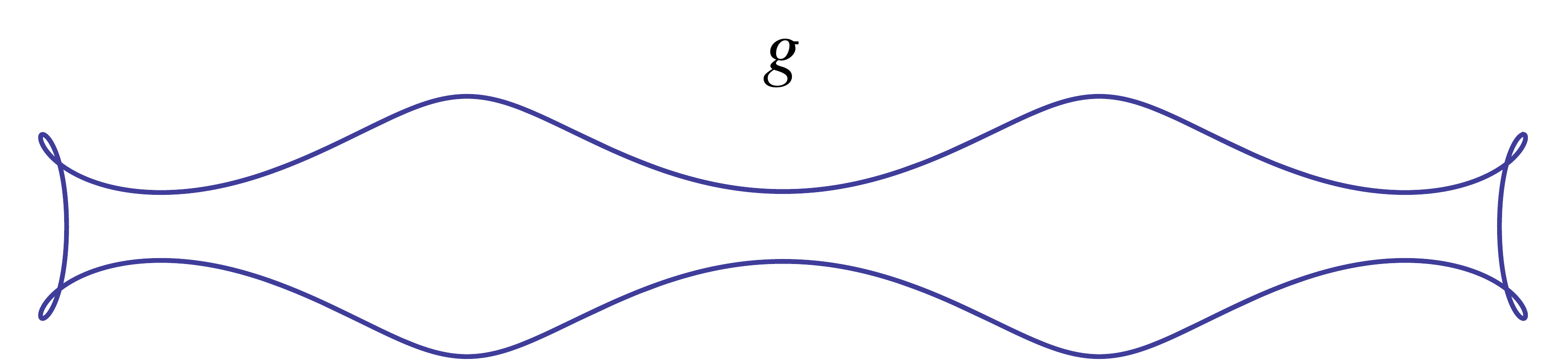}\hspace{6em}\includegraphics[scale=0.2]{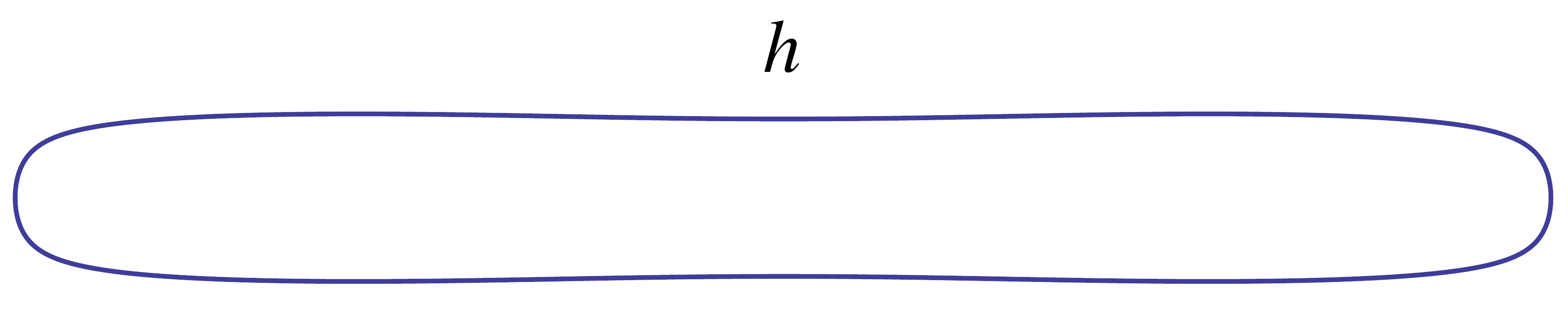}\\
\caption{Characteristic diagram of stable orbits with multiplicity 1
  for the strong bar model. Orbits belonging to the various families
  within the diagram are depicted below. Orbits with negative/positive
  $y_{0}$ are prograde/retrograde.}
\label{fig:charstrong}
\par\end{centering}
\end{figure*} 

\subsection{The Effective Potential}

\label{sec:omega}

For most galaxies, the concensus is that corotation occurs at (or just
beyond) the edge of the bar. This assertion is consistent with results
from observations \citep{Ag03}, orbital investigations \citep{Pa97},
and simulations \citep{Pf91,An92}. In our model, the following
prescription may be used to choose $\Omega_{\rm b}$ such that this
requirement is satisfied:
\begin{equation}
\Omega_{\rm b}=\Omega_{0}(a)=\sqrt{\frac{-F_{x}(a,0)}{a}},
\label{omcorot}
\end{equation}
where $F_{x}(a,0)$ is the radial force at the end of the bar long
axis. This gives the effective potential the usual structure, seen in
Fig.~\ref{fig:phieff}, where L$_1$ to L$_5$ are the familiar Lagrange
radii (see e.g., Binney \& Tremaine 1987).

If the pattern speed is large enough, however, then the well-known
morphology changes.  There exists a critical pattern speed
\begin{equation}
\Omega_{\rm crit}(a)=\big(1+a^2)^{-1/2},
\label{omcrit}
\end{equation}
such that if $\Omega_{\rm b}>\Omega_{\rm crit}$, the Lagrange point
$L_{3}$ becomes a saddle point. In addition, two more minima then
appear: $Q_{1}$ and $Q_{2}$. This effect goes unnoticed for bars where
$a\lesssim3$, since $\Omega_{\rm crit}(a)<\Omega_{0}(a)$ in this
case. Fig.~\ref{fig:effpotx} depicts the various shapes the effective
potential may take as $\Omega_{\rm b}$ is increased. In this regime,
along the $x$-axis $L_{3}$ is a maximum and $Q_{1},Q_{2}$ are
minima. Orbits can then become bound in the regions surrounding
$Q_{1}$ and $Q_{2}$, creating overdensities. The existence of
additional stationary points in the effective potential has been
noticed before by \citet{Ha00} in a numerically constructed model of
the Milky Way bar, though there the stationary points were off-axis.
Multiple Lagrange may be common in bars with flattish profiles.  The
\citet{Fr66a,Fr66b} bars, which have homogeneous density, may see
their whole major axis as a Lagrange equilibrium line when the
rotation frequency is equal to the long axis oscillation frequency.
So, flattish bars, the central parts of which may be nearly
homogeneous, may possesses multiple Lagrange points, the equilibirum
line transforming to several points.

This feature may have observational consequences.  It is well-known
that many galactic bars exhibit a peanut morphology \citep{Lu00}.
This could be produced by stars librating around such additional
on-axis Lagrange points like $Q_1$ and $Q_2$. In this picture, the
peanut morphology may be interpreted as an effect related to the
balance between centrifugal forces and self-gravity in a rotating
strong bar.

\subsection{Surfaces of Section and Characteristic Diagrams}

A powerful tool for analysing orbital structure is the Poincar\'{e}
surface of section. In a 2D potential, the phase space of a star has
four coordinates, $(x,y,\dot{x},\dot{y})$, and cannot be
visualised. However, one can see a depleted picture of this phase
space by fixing the value of $H_{\rm J}$, and then plotting points
when a given orbit intersects a chosen plane. These points are known
as consequents, and the 2D plot is a surface of section.

In this analysis, we fix $H_{\rm J}$ and then plot the values of
$(y,\dot{y})$ when an orbit upwardly intersects the $x=0$ plane
($\dot{x}>0$). We do this for a variety of initial
conditions. Periodic orbits that upwardly intersect the $x$-axis with
multiplicity $n$ appear in these diagrams as a set of $n$ invariant
points, and quasiperiodic orbits form invariant curves around the
invariant points (parent orbits). Chaotic orbits simply fill the
phase-space available to them.  The boundary of each surface of
section is the zero velocity surface (ZVS) and is defined by the
energy condition
\begin{equation}
\frac{1}{2}\dot{y}^2+\Phi_{\rm eff}(0,y)\leq\frac{1}{2}(\dot{x}^2+\dot{y}^2)+\Phi_{\rm eff}(0,y)=H_{J}(x=0).
\end{equation}

A characteristic diagram is a related concept to that of a surface of
section. A family of periodic orbits will exist at a range of
energies, and it is useful to monitor its evolution as well as
possible bifurcations. We characterise a given family by the curve
\begin{equation}
y_{0}=f(H_{j}),
\label{eq:chareq} 
\end{equation}
where $y_{0}$ is the $y$-coordinate at which the orbit upwardly
intersects the $x$-axis (i.e. using the same convention as in the
surfaces of section). A different curve of the form of
eqn~(\ref{eq:chareq}) exists for each family of periodic orbits. In
practice, one is interested in the stable orbits of low multiplicity,
since these will be the orbits that trap the most matter. Hence, in
our characteristic diagrams, only orbits of multiplicity one are
considered. In order to construct such a diagram, one needs to locate
a periodic orbit using a surface of section, and then extrapolate to
other energies by using suitable root-finding techniques
\citep{Co02,Ha93}.

\subsection{Weak and Strong Logarithmic Bars}

In order to understand the effects of bar strength, we provide results
for two cases: a weak bar and strong bar. The parameters for each
model are given in Table \ref{table:models}.  We pick $\Omega_{\rm
  b}=\Omega_{0}(a)$ in the weak bar model, and $\Omega_{\rm
  b}=\Omega_{\rm crit}(a)$ in the strong bar model. For the strong bar
case, this moves corotation to $x\simeq\pm 1.22a$.

The surfaces of section for the weak logarithmic bar all depict a very
similar structure, which is shown in Fig~\ref{fig:sos_weak}. Indicated
is the long-axis $x_1$ orbit (following the nomenclature in
\citealt{Co02}), and the corresponding quasiperiodic orbits that are
parented by it. These elongated orbits trap most of the stars in order
to form the bar. The other noticeable resonance is the retrograde $x_4$
orbit, which will be far less populated than the $x_1$ orbits, since it
is elongated along the short-axis of the bar. The surface of section
is ruled by invariant curves and there are few chaotic orbits.

For comparison, Fig.~\ref{fig:sos_strong} is a surface of section of
the strong bar. At this energy ($H_J = 1.24$), the $x_4$ family are
unstable and have disappeared from the surface of section, so the
orbital structure is dominated by $x_1$ orbits. As mentioned in
Sect~\ref{sec:omega}, if the pattern speed is large enough, two more
minima ($Q_1$ and $Q_2$) appear in the effective potential. Orbits are
indeed trapped around these minima, and this is demonstrated in
Fig.~\ref{fig:sos_extra}. This surface of section is generated using
$(x,\dot{x})$ as the phase-space coordinates and $y=0$, since
off-centre orbits do not intersect the $y$-axis. Two `islands' are
visible, with parent orbits at their centres. These are off-centre $x_4$
orbits circulating around the extra Lagrangian points $Q_{1}$ and
$Q_{2}$. Notice too that there is a clearly visible chaotic layer
associated with the separatrix.

Figs \ref{fig:charweak} and \ref{fig:charstrong} are the
characteristic diagrams for the models considered here.  The simple
nature of the orbital structure of the weak bar is clear from its
characteristic diagram, which exhibits no bifurcations or additional
families whatsoever. However, the strong bar presents a different
picture.  As the energy increases, we see successive bifurcations from
the $x_1$ sequence. These have an extended 'bow tie' shape (see orbits f
and g), albeit of more and more elaborate form. True $x_1$ orbits only
exist in the very central parts of the model (below $H_J \approx
1.25$). With their extremely slender shape, these orbits look more
like the propeller orbits identified by \cite{Ka05} in their study of
a $n=2$ Ferrers bar embedded in a Plummer sphere. However, there are
some differences. In the characteristic diagrams of \cite{Ka05}, the
$x_1$ orbits and propellers can coexist over some range of energies, with
the propeller family eventually taking over (see their Figure
1). Here, the propeller orbits bifurcate from the $x_1$ sequence. This
morphology in the characteristic diagram appears not to have been
reported before, probably because previous investigations have often
focussed on Ferrers bars. The logarithmic bar deviates very strongly
from the elliptic density contours of Ferrers bars, so qualitatively
new features in the characteristic diagram are to be expected.

We expect that these elongated and narrow orbits will trap most of the
stars in the bar region in order to form the bar backbone.  However,
even in the case of the strong bar, this potential has a relatively
simple orbital structure, with $x_1$ and then propeller orbits providing
the dominant families.

\section{Conclusions}

The paper has introduced {\it the logarithmic bar}, which is elongated
and triaxial in the central parts, but becomes axisymmetric at large
radii. It is produced by convolving a needle density with the familiar
logarithmic potential that generates a flat rotation curve~
\citep{Ev93,BT}.  Note that the entire model represents both the
luminous and dark matter in a barred galaxy, as the model retains the
appealing property of its progenitor in possessing an asymptotically
flat rotation curve. Many of the properties of the model -- the
potential, the density, the force field and the surface density along
the principal axes -- are analytic. The surface density profile along
the bar major axis is flattish, typically falling by factors of $\sim
2$ from centre to bar end. This is characteristic of bars in
early-types (SB0, SBa), for which there is ample observational
evidence for such flattish, slowly declining
profiles~\citep{El85,Se93,El96}.

Much of our insight into the orbital structure of bars is derived from
studies of Ferrers and Freeman ellipsoids~\citep[see e.g.,][]{Co02}.
These are analytically tractable, but are too homogeneous to represent
real galactic bars.  Unlike Ferrers bars, the density contours of the
logarithmic bar are not ellipsoidal, but strongly elongated and
spindly. The orbital structure of these new models is qualitatively
different from Ferrers bars.

Weak logarithmic bars have a very simple orbital structure, with
almost all stars moving on the retrograde $x_4$ and prograde $x_1$ orbits.
As the bar strength and pattern speed are increased, additional
stationary points of the effective potential can occur on the major
axis. This phenomenon appears to be new, and it allows families of
off-centered $x_4$ orbits to exist stably. This may have observational
consequences, as it is a mechanism of maintaining long-lived peanut
structures.  Strong logarithmic bars also have a simple but unusual
orbital make-up.  At low energies, the $x_1$ orbital family still exists,
although the $x_4$ family does not. However, at higher energies,
propeller orbits (essentially very narrow $x_1$ orbits with superposed
epicyclic oscillations) bifurcate from the $x_1$ sequence and take over.
Their thinness enables extremely slender bars to be built. This
provides a new family -- in addition to the one reported by
\citet{Ka05} -- for which $x_1$ orbits are not the principal means of
building the bar. The morphology of the characteristic diagrams is
therefore different to any in the literature.

Our new model has many attractive features, both for fitting to barred
spirals and for dynamical investigations. In particular, its
simplicity compares favourably with its competitors. For example,
Ferrers bars must always be supplemented by a spherical or
axisymmetric background representing halo and disk to yield a
cumbersome multi-component and multi-parameter galaxy.

Bars in late-type galaxies (SBb, SBc) are very
different~\citep{El96}. They have exponentially declining surface
density profiles along the major axis, together with Gaussian
cross-sections in the transverse directions. The versatile
\citet{Lo92} algorithm is capable of generating models with precisely
these properties. A companion paper presents the properties of these
{\it exponential bars}, which provide a fitting contrast to the
flattish logarithmic bars studied here.

\section*{Acknowledgments}

We thank David Kaufmann for his useful advice, as well as the
anonymous referee for a very helpful report. AW acknowledges the
support of STFC.

\bibliography{barredlog_paper}
\bibliographystyle{mn2e}

\end{document}